\documentclass[english]{article}
\usepackage[T1]{fontenc}
\usepackage[latin9]{inputenc}
\usepackage{color}
\usepackage{float}
\usepackage{amsmath}
\usepackage{amssymb}
\usepackage{graphicx}
\usepackage{esint}

\makeatletter

\providecommand{\tabularnewline}{\\}

\newcommand{\lyxaddress}[1]{
\par {\raggedright #1
\vspace{1.4em}
\noindent\par}
}

\usepackage{a4wide}

\makeatother

\usepackage{babel}
\begin{document}

\title{Truncated Hilbert space approach to the 2d $\phi^{4}$ theory}

\author{Z. Bajnok$^{1}$ and M. Lájer$^{2}$}

\maketitle

\lyxaddress{\begin{center}
$^{1}$MTA Lend\"ulet Holographic QFT Group, Wigner Research Centre
for Physics\\
H-1525 Budapest 114, P.O.B. 49, Hungary\\
~\\
$^{2}$ Roland Eötvös University, \\
 1117 Budapest, Pázmány Péter sétány 1/A, Hungary
\par\end{center}}
\begin{abstract}
We apply the massive analogue of the truncated conformal space approach
to study the two dimensional $\phi^{4}$ theory in finite volume.
We focus on the broken phase and determine the finite size spectrum
of the model numerically. We interpret the results in terms of the
Bethe-Yang spectrum, from which we extract the infinite volume masses
and scattering matrices for various couplings. We compare these results
against semiclassical analysis and perturbation theory. We also analyze
the critical point of the model and confirm that it is in the Ising
universality class.
\end{abstract}

\section{Introduction}

Interactive quantum field theories (QFT) cannot be solved exactly.
There are some exceptions including two dimensional integrable QFTs,
which can be investigated by the bootstrap method, but these theories
are very special, as the scattering matrix is factorized and there
is no particle production or decay. To study more realistic QFTs,
one has to rely on alternative approximative methods such as perturbation
theory, variational methods, lattice discretizations or the truncated
conformal space approach (TCSA) just to mention a few.

The truncated conformal space approach was originally developed by
Yurov and Zamolodchikov to study the relevant perturbation of the
Lee-Yang conformal field theory \cite{Yurov:1989yu}. The idea of
the method is to put the system in a finite volume, which makes the
spectrum discrete, then one can truncate the Hilbert space by keeping
states below some energy cut and diagonalize the truncated Hamiltonian
numerically to get an approximate spectrum as the function of the
volume. This is basically the usual variational method of quantum
mechanics with the unperturbed energy levels as basis states and the
obtained results are not perturbative. 

The TCSA method has been extended to analyze the perturbation of other
conformal field theories \cite{Feverati:1998va,Beria:2013hz,Konik:2015bia},
too. The advantage of the method is that it can be applied for non-integrable
perturbations \cite{Bajnok:2000ar}. Indeed it has been successfully
applied to describe such phenomena as phase transitions, decay rates,
form factors or boundary entropies \cite{Bajnok:2000ar,Pozsgay:2006wb,Pozsgay:2007kn,Takacs:2011aa}.
It also has a natural extension for models with boundaries and with
a defect \cite{Bajnok:2001py,Bajnok:2013waa}. Recently the method
has been extended even to higher dimensional theories \cite{Hogervorst:2014rta}. 

The simplest non-integrable interacting quantum field theory is probably
the two dimensional $\phi^{4}$ theory. The TCSA method was used in
\cite{Coser:2014lla} to study the spectrum by compactifying the zero
mode of the field. This approach gives a good result for small couplings.
Alternatively, in \cite{Rychkov:2014eea,Rychkov:2015vap} the authors
used the finite volume massive basis to build up the truncated Hilbert
space and analyzed both the unbroken and the broken phases of the
model for a wide range of couplings. There special emphasis was put
on developing the method and on analyzing carefully the convergence
properties in eliminating the truncation. The authors also searched
for the critical point, beyond which the parity symmetry is broken
and checked Chang duality, by which one can realize the same system
with two different parameter sets, which are related to each other
in a strong-weak fashion. The truncation dependence of the spectrum
was further analyzed in \cite{Elias-Miro:2015bqk}. The aim of our
paper is to use the same truncation of the finite volume massive basis
for the broken phase and obtain results for a wide range of couplings
similarly to \cite{Rychkov:2014eea,Rychkov:2015vap}. We focus on
the volume dependence of the spectrum and try to extract the infinite
volume parameters from finite size effects. This analysis requires
precise data, which we achieved, first, by implementing an efficient
diagonalizing routine for large sparse matrices, and second, by an
extrapolation in the truncation level. We compare the extrapolated
finite size spectrum with semiclasssical results, perturbative results
and also with the Bethe-Yang (BY) spectrum. This BY spectrum incorporates
the leading finite size corrections based on the infinite volume masses
and scattering data and enables one to determine these infinite volume
characteristics at moderate volumes when the truncation errors can
be controlled.

The paper is organized as follows:

In section 2 we review the classical and semiclassical results. Section
3 contains the definition of the quantum model. In section 4 we explain
our truncated Hilbert space method. Theoretical predictions are summarized
in section 5, while numerical results are presented in section 6 and
compared to both the semiclassical, the perturbative and to the Bethe-Yang
spectrum. From this comparison we extract the infinite volume mass
spectrum and scattering phases. We also analyze the spectrum near
the critical point. We close the paper with conclusions in section
7. Appendix A explains volume dependent normal ordering and its relation
to the ground-state energy. Appendix B introduces the parametrization
of the scattering matrix from unitarity relations, while Appendix
C explains the numerical implementations.

\section{Classical, semiclassical and perturbative results}

In this section we recall the available results for our theory
\begin{equation}
\mathcal{L=}\frac{1}{2}\left(\partial_{t}\phi\right)^{2}-\frac{1}{2}\left(\partial_{x}\phi\right)^{2}+\frac{1}{2}m^{2}\phi^{2}-\frac{\lambda}{4!}\phi^{4}
\end{equation}
We follow \cite{Coser:2014lla}, but see also \cite{Rajaraman:1982is,Dashen:1974cj}
for more details. Sometimes we simplify formulas by introducing $q$
as $\lambda=6q^{2}$. We take $\hbar=c=1$. As in two dimensions the
scalar field is dimensionless, both $m$ and $q$ have dimension of
energy, i.e. $1$. The parameter $m$ can be chosen to determine the
mass scale and then the physics depends on the dimensionless ratio:
$\frac{m}{q}$. Observe that the coefficient of $\phi^{2}$ has a
``wrong'' sign, such that we analyze the theory when the $\mathbb{Z}_{2}$
parity symmetry will be spontanously broken.

\subsection{Classical analysis}

The potential has two minima at $\pm\frac{m}{q}$ and its expansions
around the minima reads as 
\begin{equation}
V\left(\pm\frac{m}{q}+\eta\right)=-\frac{m^{4}}{4q^{2}}+m^{2}\eta^{2}\pm mq\eta^{3}+\frac{q^{2}}{4}\eta^{4}\equiv V_{0}+\frac{m_{0}^{2}}{2}\eta^{2}+\dots\label{eq:pertdef}
\end{equation}
The mass of the small fluctuations is $m_{0}=\sqrt{2}m$ and the minimum
of the potential is $V_{0}=-\frac{m^{4}}{4q^{2}}$. The two static
solutions, called the kink and the antikink, interpolate between the
vacua as
\begin{equation}
\phi_{\pm}=\pm\frac{m}{q}\tanh\left(\frac{mx}{\sqrt{2}}\right)
\end{equation}
and have masses 
\begin{equation}
M_{0}=\frac{2\sqrt{2}m^{3}}{3q^{2}}=\frac{m_{0}}{g}\quad;\qquad g=\frac{3q^{2}}{2m^{2}}=\frac{\lambda}{4m^{2}}
\end{equation}
The breather-like bound state of the kink and the antikink is not
a stable solution as it radiates very slowly \cite{Fodor:2008es}.

\subsection{Semiclassical considerations}

We can quantize the theory semiclassically. For this normal ordering
is needed, which has to be defined wrt. a free theory. We recall for
later reference that in the calculations \cite{Rajaraman:1982is,Dashen:1974cj}
the reference theory was chosen to be the infinite volume free massive
theory describing the fluctuations around any of the minima with mass
$m_{0}$. The relevant quantities at the quantum level are the masses
and the scattering matrix.

The semiclassical corrections to the kink masses read as
\begin{equation}
M=M_{0}-m_{0}c+\mathcal{O}\left(q^{2}\right)\quad;\qquad c=\frac{3}{2\pi}-\frac{1}{4\sqrt{3}}
\end{equation}
where, as we pointed out, the parameters $m,q$ are understood as
the coefficients of the normal ordered perturbations wrt. $m_{0}$.

The breathers can be considered as kink-antikink bound states with
masses \cite{Mussardo:2006iv}
\begin{equation}
m_{n}=2M\sin\Bigl(n\frac{\zeta}{2}\Bigr)
\end{equation}
where 
\begin{equation}
\zeta=\frac{g}{1-gc}=\frac{\frac{3q^{2}}{2m^{2}}}{1-\frac{3q^{2}}{2m^{2}}\left(\frac{3}{2\pi}-\frac{1}{4\sqrt{3}}\right)}\label{eq:semizeta}
\end{equation}
Although classically there is no stable breather solution due to the
radiation into the small fluctuations of the field, at the quantum
level the small fluctuations are indentified with the breather itself
thus can be stable. Higher breathers are bound states of the lightest
one
\begin{equation}
m_{n}=m_{0}n\left(1-\frac{3q^{4}}{32m^{4}}n^{2}+\dots\right)=m_{0}n\left(1-\frac{g^{2}}{24}n^{2}+\dots\right)\label{eq:semimn}
\end{equation}
The existence of the $n^{th}$ bound state requires the scattering
pole $n\zeta$ to be in the physical strip: $n\zeta<\pi$, i.e. $\zeta<\frac{\pi}{n}$.
The stability against the lighest particle further requires that $m_{n}<2m_{1}$,
which is always satisfied for $m_{2}$ but never for $m_{3}$ and
for the higher ones. It was conjectured in \cite{Mussardo:2006iv}
that this picture survives at the quantum level in the sense that
we cannot have more than two breathers. Let us anticipate that in
the deep quantum domain even the first breather is unstable against
decaying into a kink and antikink pair. 

The scattering matrix for the lightest breather satisfies unitarity
and crossing symmetry below the two-particle threshold
\begin{equation}
S(\theta)=S(-\theta)^{-1}=S(i\pi-\theta)\label{eq:SUC}
\end{equation}
Here we introduced the rapidity variable, see Appendix B for details.
The simplest non-trivial solution of (\ref{eq:SUC}) is 
\begin{equation}
S(\theta,i\kappa)=\frac{\sinh\theta+i\sin\kappa}{\sinh\theta-i\sin\kappa}
\end{equation}
It implies a bound state with mass $m_{2}=2m_{1}\cos\frac{\kappa}{2}$.
Comparing this to the semiclassical mass formula \ref{eq:semimn},
we conclude that $\kappa=\zeta$. 

From the analysis of the small fluctuations over the kink solution
the classical limit of the breather-kink scattering matrix can also
be extracted \cite{Rajaraman:1982is}: 
\begin{equation}
S_{bk}(\theta)=S\left(\theta,i\frac{\pi}{2}\right)S\left(\theta,i\frac{\pi}{6}\right)\label{eq:Skbclass}
\end{equation}
The pole at $\theta=i\pi/2$ is related to the translational mode,
while the pole at $\theta=i\pi/6$ signals an excited kink state with
mass 
\begin{equation}
M^{*}=M+m_{0}\frac{\sqrt{3}}{2}
\end{equation}

\subsection{Related integrable models and perturbative results}

There are two integrable models, which are related to our theory.
The sinh-Gordon theory is defined by the Lagrangian

\begin{equation}
\mathcal{L}=\frac{1}{2}\left(\partial_{t}\phi\right)^{2}-\frac{1}{2}\left(\partial_{x}\phi\right)^{2}-\frac{m^{2}}{b^{2}}(\cosh b\phi-1)
\end{equation}
It contains only one type of particle with scattering matrix
\begin{equation}
S(\theta)=S(\theta,-i\pi p)\quad;\qquad p=\frac{b^{2}}{8\pi+b^{2}}=\frac{b^{2}}{8\pi}+\dots
\end{equation}
The sine-Gordon theory can be obtained by an analytical continuation
$b\to i\beta$, which admits kink and anti-kink solutions, too. 

The other interesting theory is the Bullough-Dodd theory 
\begin{equation}
\mathcal{L}=\frac{1}{2}\left(\partial_{t}\phi\right)^{2}-\frac{1}{2}\left(\partial_{x}\phi\right)^{2}-\frac{m^{2}}{6g^{2}}\left(2e^{g\phi}+e^{-2g\phi}\right)
\end{equation}
Its scattering matrix has the form \cite{Arinshtein:1979pb} 
\begin{equation}
S(\theta)=S\left(\theta,i\frac{\pi}{3}\right)S\left(\theta,i\frac{\pi B}{3}\right)S\left(\theta,i\frac{\pi(2-B)}{3}\right)\,\quad;\qquad B=\frac{g^{2}}{2\pi}\frac{1}{1+\frac{g^{2}}{4\pi}}=\frac{g^{2}}{2\pi}+\dots\label{eq:BD_S-matrix}
\end{equation}
Combining the perturbative expansions of the two models one can extract
the leading order perturbative S-matrix of our model: 
\begin{equation}
S(\theta)=1+2ig\frac{(2+\cosh2\theta)}{\sinh3\theta}+\dots\label{eq:pertS}
\end{equation}
Direct perturbative calculation for the mass correction gives \cite{Rychkov:2015vap}:
\begin{equation}
m_{pert}^{2}(g)-m_{0}^{2}=m_{0}^{2}\left(-\frac{g}{2\sqrt{3}}-0.0317182g^{2}\right)\label{mpert}
\end{equation}

We would like to compare all of the results of this section against
the finite size spectrum we obtain from our truncated Hilbert space
approach.

\section{Quantum theory}

At the quantum level we analyze the spectrum of the Hamiltonian in
a finite volume $L$:
\begin{equation}
H=\int_{0}^{L}\mathcal{H}\, dx\quad;\qquad\mathcal{H}=\frac{1}{2}\left(\partial_{t}\phi\right)^{2}+\frac{1}{2}\left(\partial_{x}\phi\right)^{2}-\frac{1}{2}m^{2}\phi^{2}+\frac{\lambda}{4!}\phi^{4}
\end{equation}
where $\phi(x)=\pm\phi(x+L)$, i.e. we consider both periodic ($+$)
and antiperiodic ($-$) boundary conditions in order to accommodate
one-particle kink states. 

As operators are multiplied in the same position we need to regulate
them by normal ordering. Normal ordering is performed with respect
to some prescribed mass scale $\mu$ and volume $L$, thus the Hamiltonian
density takes the form:
\begin{equation}
\mathcal{H}=\mathcal{H}_{0}^{\mu,L}+g_{2}(\mu,L):\!\phi^{2}\!:_{\mu,L}+g_{4}(\mu,L):\!\phi^{4}\!:_{\mu,L}+g_{0}(\mu,L)\label{eq:Hmu}
\end{equation}
Here the field is expressed in terms of creation-annihilation operators
as 
\begin{equation}
\phi(x,t)=\sum_{n}\frac{1}{\sqrt{2\omega_{n}L}}\left(a_{n}e^{-i\omega_{n}t+ik_{n}x}+a_{n}^{+}e^{i\omega_{n}t-ik_{n}x}\right)
\end{equation}
where 
\begin{equation}
[a_{n},a_{m}]=[a_{n}^{+},a_{m}^{+}]=0\quad;\qquad[a_{n},a_{m}^{+}]=\delta_{n,m}
\end{equation}
and 
\begin{equation}
\omega_{n}=\sqrt{\mu^{2}+k_{n}^{2}}\quad;\qquad k_{n}=\frac{2\pi}{L}n
\end{equation}
The index $n$ takes integer values for periodic, while half-integer
values for antiperiodic boundary conditions. Normal ordering, denoted
above by $::_{\mu,L}$, is understood that all creation operators
$a_{n}^{+}$ are on the left of the annihilation operators $a_{n}$.
The vacuum is chosen to be annihilated by all $a_{n}$. The free Hamiltonian
is 
\begin{equation}
H_{0}^{\mu,L}=\frac{1}{2}\int_{0}^{L}:\left(\partial_{t}\phi\right)^{2}+\left(\partial_{x}\phi\right)^{2}+\mu^{2}\phi^{2}:_{\mu,L}+E_{0}^{\pm}(\mu,L)=\sum_{n}\omega_{n}\, a_{n}^{+}a_{n}+E_{0}^{\pm}(\mu,L)
\end{equation}
where the renormalized ground-state energy is chosen to be
\begin{equation}
E_{0}^{\pm}(\mu,L)=\mu\int_{-\infty}^{\infty}\frac{d\theta}{2\pi}\cosh\theta\log\left(1\mp e^{-mL\cosh\theta}\right)
\end{equation}
for periodic/antiperiodic boundary conditions. Since both the ground-state
energy and the ground-state energy density is infinite for the non-renormalized
Hamiltonian, we have to introduce proper renormalization. In Appendix
A, we show how $E_{0}^{\pm}(\mu,L)$ appears from comparing different
renormalization schemes. In our scheme the vacuum energy density is
normalized to zero and the ground-state energy is chosen to vanish
for infinite volume. 

To relate (\ref{eq:Hmu}) to another normal ordering at scale $\mu'$
and volume $L'$, we recall that \cite{Coser:2014lla,Rychkov:2014eea}
\begin{equation}
:\!\phi^{2}\!:_{\mu',L'}=:\!\phi^{2}\!:_{\mu,L}+\Delta Z\quad;\quad\Delta Z=Z(\mu,L)-Z(\mu',L')
\end{equation}
where 
\begin{equation}
Z(\mu,L)=\sum_{\vert n\vert<N}\frac{1}{2\omega_{n}L}
\end{equation}
$Z(\mu,L)$ itself is only meaningful with a UV cutoff $N$, but in
the difference, $\Delta Z$, the limit $N\rightarrow\infty$ can be
taken. Similarly 
\begin{equation}
:\!\phi^{4}\!:_{\mu',L'}=:\!\phi^{4}\!:_{\mu,L}+6\Delta Z:\!\phi^{2}\!:_{\mu,L}+3(\Delta Z)^{2}
\end{equation}
As we already mentioned, the scheme in which the semiclassical calculations
were done is defined by normal ordering at $L'=\infty$ and $\mu'=m_{0}$.
The parameters are chosen there as 
\begin{equation}
g_{2}\left(m_{0},\infty\right)=-\frac{3}{2}m^{2}\quad;\qquad g_{4}\left(m_{0},\infty\right)=\frac{q^{2}}{4}\quad;\qquad g_{0}\left(m_{0},\infty\right)=0
\end{equation}
To relate this scheme to the one with mass $m$ and $L=\infty$, we
calculate 
\begin{equation}
Z(m,\infty)-Z(m_{0},\infty)=\frac{1}{2\pi}\int_{0}^{\infty}\left(\frac{1}{\sqrt{1+k^{2}}}-\frac{1}{\sqrt{2+k^{2}}}\right)dk=\frac{\ln2}{4\pi}
\end{equation}
 such that the coefficients of the same physical system in this scheme
are 
\begin{equation}
g_{2}(m,\infty)=-m^{2}+\frac{3q^{2}\ln2}{8\pi}=m^{2}\left(g\frac{\ln2}{4\pi}-1\right)\quad;\qquad g_{4}(m,\infty)=\frac{q^{2}}{4}
\end{equation}
and 
\begin{equation}
g_{0}(m,\infty)=-\frac{3}{2}m^{2}\frac{\ln2}{4\pi}+\frac{3}{4}q^{2}\left(\frac{\ln2}{4\pi}\right)^{2}
\end{equation}
As we work in the finite volume scheme we introduce \cite{Rychkov:2014eea}:
\begin{equation}
z(mL)=Z(m,L)-Z(m,\infty)
\end{equation}
For periodic boundary condition we have 
\begin{equation}
z^{+}(mL)=\int_{0}^{\infty}\frac{d\theta}{\pi}\left(e^{mL\cosh\theta}-1\right)^{-1}
\end{equation}
while for antiperiodic boundary condition we found
\begin{equation}
z^{-}(mL)=2z^{+}(2mL)-z^{+}(mL)
\end{equation}
The coefficient at volume $L$ of our theory are given by: 
\begin{eqnarray}
g_{2}(m,L) & = & m^{2}\tilde{g}_{2}(mL)\quad;\qquad\tilde{g}_{2}(mL)=\left[g\,\tilde{z}(mL)-1\right]\quad;\quad\tilde{z}(x)=z^{\pm}(x)+\frac{\ln2}{4\pi}\nonumber \\
g_{4}(m,L) & = & m^{2}\tilde{g}_{4}\quad;\qquad\tilde{g}_{4}=\frac{g}{6}\label{eq:g2g4}
\end{eqnarray}
The correction to the vacuum energy density due to normal ordering
is 
\begin{equation}
g_{0}(m,L)=m^{2}\tilde{g}_{0}(m,L)\quad;\qquad\tilde{g}_{0}(m,L)=-z^{\pm}(mL)-\frac{3}{2}\frac{\ln2}{4\pi}+\frac{g}{2}\tilde{z}(mL)^{2}
\end{equation}
Since $z^{\pm}(L)$ is exponentially small, it can be often neglected. 

It is tempting to introduce the normal ordering at $\mu=0$ and $L=\infty$,
as then the scale would not appear as a dimensionful parameter \cite{Coser:2014lla}.
Choosing $\mu=0$ implies, however, that $Z(0,\infty)-Z(m_{0},\infty)$
is IR divergent, which can be regulated by an IR cutoff $a,$ 
\begin{equation}
Z(0,\infty)=\frac{1}{2\pi}\int_{a}^{\Lambda}\frac{dk}{k}=\frac{1}{2\pi}\ln\left(\frac{\Lambda}{a}\right)
\end{equation}
but then the theory will depend on the regulator $a$. Naively we
could just use $Z(0,L)$ (conformal scheme) instead of introducing
$Z(0,\infty)$ but we should keep in mind that $L\to\infty$ is not
a safe limit as it basically puts $a$ to zero. As a result, the mass
of the breather will diverge logarithmically in the $L\to\infty$
limit in this conformal scheme.

\section{Truncated Hilbert space approach}

In this section we explain how we diagonalize the finite volume Hamiltonian.
For numerical investigations we introduce dimensionless quantities
\begin{equation}
\frac{H}{m}=h\quad;\qquad mL=l\quad;\quad\omega_{n}=m\tilde{\omega}_{n}=m\sqrt{1+\frac{4\pi^{2}n^{2}}{l^{2}}}
\end{equation}
such that the dimensionless Hamiltonian can be written as 
\begin{equation}
\tilde{h}=\tilde{h}_{0}+\int_{0}^{L}mdx\left[\tilde{g}_{2}(l):\!\phi^{2}\!:+\tilde{g}_{4}:\!\phi^{4}\!:+\tilde{g}_{0}(l)\right]
\end{equation}
where 
\begin{equation}
\tilde{h}_{0}=\sum_{n}\tilde{\omega}_{n}(l)a_{n}^{+}a_{n}+\tilde{e}_{0}^{\pm}(l)
\end{equation}
and $\tilde{e}_{0}^{\pm}(l)=E_{0}^{\pm}(mL)/m$. From now on all normal
orderings are evaluated at scale $m$ and volume $L$, unless otherwise
indicated. 

Technically we use the Hilbert space built up by the oscillators of
the free Hamiltonian in volume $L$ and diagonalize $h$ at a certain
truncation level. Clearly we have two dimensionless parameters: the
dimensionless volume $l$ and the quartic coupling parameter $g$.
The dimensionful mass $m$ sets purely the energy scale.

\subsection{Mini Hilbert space approach}

For periodic boundary condition we found it useful to adapt the ``mini-superspace''
approach. This is the method which was used many times in analyzing
the Liouville theory and deals with the zero mode of the field separately
\cite{Seiberg:1990eb,Zamolodchikov:1995aa,Bajnok:2007ep}. The same
approach was used also in \cite{Rychkov:2015vap} to handle the zero
mode. Here we also found that the IR mode generated by $a_{0}$ and
$a_{0}^{+}$ plays a very special role in forming the low lying states
in the spectrum. We thus decompose the elementary field as 
\begin{equation}
\phi(x,0)=\tilde{\phi}(x,0)+\phi_{0}\quad;\qquad\phi_{0}=\frac{1}{\sqrt{2l}}\left(a_{0}+a_{0}^{+}\right)
\end{equation}
and analyze the mode $\phi_{0}$ first. It has the Hamiltonian 
\begin{equation}
h^{\textrm{mini}}=\tilde{\omega}_{0}\, a_{0}^{+}a_{0}+\frac{1}{2}\tilde{g}_{2}(l):\!\left(a_{0}+a_{0}^{+}\right)^{2}\!:+\frac{1}{4l}\tilde{g}_{4}:\!\left(a_{0}+a_{0}^{+}\right)^{4}\!:\label{eq:hmini}
\end{equation}
which we diagonalize seperately and use its eigenvalues and eigenvectors
to calculate the matrix elements of the full Hamiltonian:
\begin{equation}
h=h^{\textrm{mini}}+\tilde{h}_{0}+\int_{0}^{L}mdx\left[\tilde{g}_{0}(l)+\tilde{g}_{2}(l):\!\tilde{\phi}^{2}:+\tilde{g}_{4}\left(:\!\tilde{\phi}^{4}\!:+4:\!\tilde{\phi}^{3}\!:\phi_{0}+6:\!\tilde{\phi}^{2}\!:\phi_{0}^{2}\right)\right]\label{eq:hfull}
\end{equation}
where $\tilde{h}_{0}=h_{0}-h_{0}^{\mathrm{mini}}$ and $h_{0}^{\mathrm{mini}}=\tilde{\omega}_{0}\, a_{0}^{+}a_{0}$
denotes the harmonic oscillator Hamiltonian of the zero mode.

\subsection{Technical implementations}

For periodic boundary condition we split the Hilbert space to the
mini-Hilbert space containing the zero mode of the field and to the
rest: 
\begin{equation}
\mathcal{H}=\mathcal{H}^{\mathrm{mini}}\otimes\tilde{\mathcal{H}}\label{eq:hspace}
\end{equation}
The mini-Hilbert space $\mathcal{H}^{\mathrm{mini}}$ is generated
from the vacuum by acting with $a_{0}^{+}$, while $\tilde{\mathcal{H}}$
by acting with all other creation operators $a_{n\neq0}^{+}$. Technically
we determine the spectrum of $h^{\mathrm{mini}}$ first. For this
we choose $2000$ eigenstates of $h_{0}^{\mathrm{mini}}$ and diagonalize
$h^{\mathrm{mini}}$ on these states. We found it enough to use only
the first $6$ states to build up the relevant product basis for (\ref{eq:hspace}). 

For antiperiodic boundary condition we do not have the analogue of
the mini Hilbert space, so we diagonalize the Hamiltonian on the full
truncated Hilbert space. 

We want to keep the dimension of the truncated space for each volume.
Since by changing the volume the free energy levels are changing,
too, we decided to generate the basis for the Hilbert space at $mL=1$.
At such a small volume the spectrum is close to the conformal spectrum
and the energy can be approximately measured in units of $\frac{2\pi}{mL}$.
Consequently, the mode $(a_{n}^{+})^{j_{n}}$ contributes to this
``conformal energy'' as $j_{n}\vert n\vert.$ We introduce integer
cuts in this unit, which we increase two by two (one by one in the
antiperiodic sector), usually between $14$ and $26$. With each cut
we keep the same oscillator content for any dimensionless volume,
which typically takes integer values up to $l=30$. 

There are conserved charges in the theory, which help to decrease
the dimension of the Hamiltonian below the energy cut in each sectors.
Momentum is a conserved quantity and we focus through the whole paper
on the zero-momentum sector. Field parity $\phi\leftrightarrow-\phi$
is also a symmetry of the full theory. As a consequence we can diagonalize
the Hamiltonian separately on the even and odd particle number sectors.
Additionally, there is also space parity $x\leftrightarrow-x$, but
in this paper we will not exploit this symmetry.

\section{Description of the spectrum}

According to Lüscher, the finite size spectrum of a quantum field
theory can be described in terms of the infinite volume data \cite{Luscher:1985dn,Luscher:1986pf}.
These data consist of the masses of the excitations, their dispersion
relation and the various scattering matrices. The leading finite size
correction comes from momentum quantization, which is described by
Bethe-Yang type equations and expresses the periodicity of the multiparticle
wave function\cite{Luscher:1986pf}. These corrections are polynomial
in the inverse power of the volume. There are exponentially small
corrections, too, and they are related to vacuum polarization effects,
or some other words, virtual particles \cite{Luscher:1985dn,Bajnok:2008bm}.

\subsection{Infinite volume data}

In infinite volume, the theory is parametrized by the masses of the
excitations and their scattering matrices. We use unitarity and consistency
with a generalized bootstrap idea to restrict the form of the scattering
matrices and masses. The bootstrap idea states that the scattering
matrix of the bound state is the product of the scattering matrices
of it constituents. It is definitely satisfied in integrable theories
and we expect that it is also satisfied for nonintegrable theories
below the inelastic threshold. 

Let us start with the breathers. According to the semiclassical analysis,
we can have at most two breathers. Let us denote the lighter one by
$B_{1}$ and its mass by $m_{1}$, while the heavier is denoted by
$B_{2}$, whose mass is $m_{2}$. As we are in a relativistic theory
their dispersion relation is 
\begin{equation}
E_{i}(p)=\sqrt{m_{i}^{2}+p^{2}}
\end{equation}
This motivates to introduce the rapidity parameter:
\begin{equation}
E_{i}(\theta)=m_{i}\cosh\theta\quad;\qquad p_{i}(\theta)=m_{i}\sinh\theta
\end{equation}
Lorentz transformation simply shifts the rapidity and, as a consequence,
the scattering matrix depends only on their differences. Since $B_{2}$
is the bound state of two $B_{1}$s, the $B_{1}B_{1}\to B_{1}B_{1}$
two-particle scattering matrix, which we denote simply by $S_{11}(\theta)$,
must have a pole whenever $B_{2}$ is in the spectrum. In Appendix
B we derive a parametrization of the scattering matrix with the existence
of the required pole from unitarity: 
\begin{equation}
S_{11}(\theta)=S(\theta,i\alpha)S_{11}^{\mathrm{exp}}(\theta)\label{eq:S11}
\end{equation}
where 
\begin{equation}
S_{11}^{\mathrm{exp}}(\theta)=\exp\left\{ -\int_{\theta_{t}}^{\infty}\frac{d\theta'}{2\pi i}\rho_{11}(\theta')\log S(\theta',\theta)\right\} \quad;\qquad S(\theta,x)=\frac{\sinh\theta+\sinh x}{\sinh\theta-\sinh x}
\end{equation}
and the position of the pole $i\alpha$ and the density $\rho_{11}(\theta)$
is an unknown function of the dimensionless parameter $g$. The threshold
rapidity $\theta_{t}$ corresponds to the kinematical point where
the $B_{1}B_{1}\to B_{1}B_{2}$ or the $B_{1}B_{1}\rightarrow\overline{K}K$
channels opens. Semiclassically, for small $g$, according to \ref{eq:semizeta},
the parameter $\alpha$ starts as 
\begin{equation}
\alpha=g+\dots\label{eq:semialpha}
\end{equation}
In the analysis we will assume that $\theta_{t}$ is large and the
exponential piece can be neglected. It would be very interesting,
although not easy, to measure this piece and compare it to other scattering
matrix elements.

Since the scattering matrix has a pole at $i\alpha$: 
\begin{equation}
S_{11}(\theta)=i\frac{\Gamma^{2}}{\theta-i\alpha}\quad;\qquad\Gamma=\sqrt{2\tan\alpha}
\end{equation}
the mass of the bound state $m_{2}$ can be expressed as 
\begin{equation}
m_{2}=2m_{1}\cos\frac{\alpha}{2}\label{eq:b2mass}
\end{equation}
Below the $\theta_{t}$ threshold the scattering matrix behaves as
if it were in an integrable theory. As a consequence one can try to
use bootstrap to calculate the scattering matrix of $B_{2}$ on $B_{1}$
\begin{equation}
S_{21}(\theta)=S_{11}\left(\theta-i\frac{\alpha}{2}\right)S_{11}\left(\theta+i\frac{\alpha}{2}\right)\tilde{S}_{21}^{\mathrm{exp}}(\theta)=S\left(\theta,i\frac{\alpha}{2}\right)S\left(\theta,i\frac{3\alpha}{2}\right)\tilde{S}_{21}^{\mathrm{exp}}(\theta)
\end{equation}
however, this scattering could be intercepted by the $B_{2}B_{1}\to B_{1}B_{1}$
process, which is taken into account partially by $\tilde{S}_{21}^{\mathrm{exp}}(\theta)$.
Additionally the appearing pole at $i\frac{3\alpha}{2}$ should correspond
to a resonance, since $B_{3}$ is not a stable particle, i.e. it should
not lie on the imaginary line. All together the correct parametrization
possibly is
\begin{equation}
S_{21}(\theta)=S\left(\theta,i\frac{\alpha}{2}\right)S_{21}^{\mathrm{exp}}(\theta)
\end{equation}

\begin{figure}[H]
\centering{}\includegraphics[width=4cm]{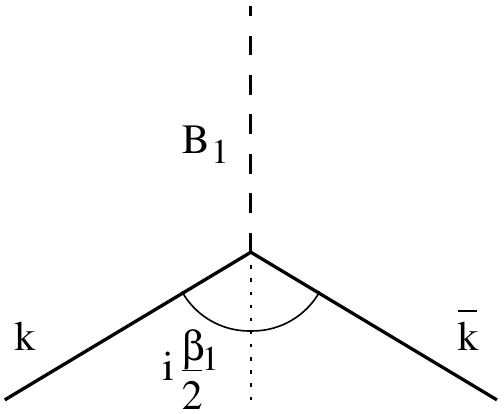} \hspace{2cm}\includegraphics[width=5cm]{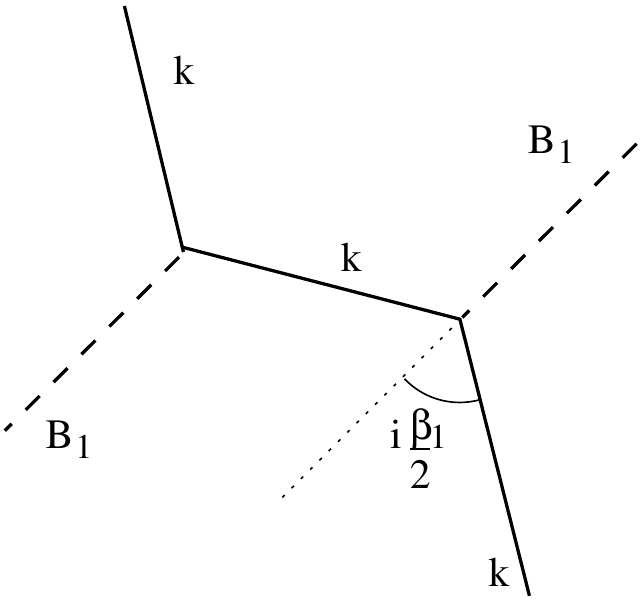}\protect\caption{The kink--anti-kink $B_{1}$ fusion process, together with the corresponding
pole in the kink $B_{1}$ scattering process.}
\label{kinkB1}
\end{figure}

Let us focus on the kink and the antikink now. The $\mathbb{Z}_{2}$
symmetry implies that these particles are degenerate in their mass,
what we denote by $M$. Interestingly the kink--antikink scattering
matrix is also diagonal and satisfies crossing symmetry and unitarity
below the threshold \cite{Mussardo:2006iv}:
\begin{equation}
S_{k\bar{k}}(\theta)=S_{\bar{k}k}(i\pi-\theta)=S_{\bar{k}k}(-\theta)^{-1}=S_{\bar{k}k}(\theta)
\end{equation}
where in the last line we used charge conjugation symmetry. A kink--antikink
pair can fuse either to $B_{1}$ or to $B_{2}$. The corresponding
angles are $\beta_{1}$ and $\beta_{2}$, such that the masses are
$m_{n}=2M\cos\frac{\beta_{n}}{2}$. The fusion process is shown on
Figure \ref{kinkB1}. This further implies that 
\begin{equation}
S_{k\bar{k}}(\theta)=-S\left(\theta,i\beta_{1}\right)S\left(\theta,i\beta_{2}\right)S_{k\bar{k}}^{\mathrm{exp}}(\theta)
\end{equation}
where we included $S_{k\bar{k}}^{\mathrm{exp}}(\theta)$ for later
consistency, which has a similar form as (\ref{eq:S11}). Here it
is assumed that we are in the domain where the kink is the lightest
particle. Note that kinematically it happens in a different domain
of the coupling $g$ , where we can trust in (\ref{eq:S11}) without
the exponential piece. 

The kink-$B_{1}$ sector has a topological charge which separates
it from the neutral sector. Thus, even for $m_{1}<M$, the scattering
matrix $S_{k1}$ also satisfies unitarity and crossing symmetry below
the kink-$B_{2}$ threshold 
\begin{equation}
S_{k1}(\theta)=S_{1\bar{k}}(i\pi-\theta)=S_{1k}(-\theta)^{-1}=S_{1k}(\theta)=S_{1\bar{k}}(\theta)
\end{equation}
 where again we used parity and charge conjugation symmetry. Due to
the kink--anti-kink fusion process this scattering matrix must have
a pole at $i\frac{\beta_{1}}{2}$ as demonstrated on Figure \ref{kinkB1}.
Taking into account that in the classical limit (\ref{eq:Skbclass})
we have a pole at $\theta=i\pi/6$ , which could get quantum correction
$\gamma$, we arrive at the parametrization 
\begin{equation}
S_{k1}(\theta)=-S\left(\theta,i\frac{\beta_{1}}{2}\right)S\left(\theta,i\left(\frac{\pi}{6}+\gamma\right)\right)S_{k1}^{\mathrm{exp}}(\theta)
\end{equation}
Demanding that fusing a kink with an antikink we obtain the breather-breather
S-matrix 
\begin{equation}
S_{k1}\left(\theta+i\frac{\beta_{1}}{2}\right)S_{\bar{k}1}\left(\theta-i\frac{\beta_{1}}{2}\right)=S_{11}(\theta)
\end{equation}
the existence of the factor $S(\theta,i\alpha)$implies that $\beta_{1}=\pi-\alpha$
(for $\alpha<\frac{\pi}{2}$). This actually gives the expected mass
relation $m_{1}=2M\sin\frac{\alpha}{2}$ and is also consistent with
the fact that in the classical limit (\ref{eq:Skbclass}) $\beta_{1}/2$
should be $\pi/2$. 

Interestingly the fusion also gives two additional factors 
\begin{equation}
S_{11}(\theta)\sim S(\theta,i\alpha)S\left(\theta,i\left(\frac{2\pi}{3}-\frac{\alpha}{2}+\gamma\right)\right)S\left(\theta,i\left(\frac{4\pi}{3}-\frac{\alpha}{2}-\gamma\right)\right)\tilde{S}_{11}^{\mathrm{exp}}(\theta)
\end{equation}
which indicates a pole in the physical strip at $\theta=\frac{2\pi}{3}-\frac{\alpha}{2}+\gamma$.
As this pole is already above the $\theta_{t}$ threshold we shouldn't
take it very seriously. Nevertheless, if we would take it, the only
interpretation could be a self-fusing pole, which is also forced by
having a cubic vertex in the perturbative definition (\ref{eq:pertdef}).
This implies the relation $\gamma=\frac{\alpha}{2}$ and leads to
the scattering matrix: 
\begin{equation}
S_{11}(\theta)\sim S(\theta,i\alpha)S\left(\theta,i\frac{2\pi}{3}\right)S\left(\theta,i\left(\frac{4\pi}{3}-\alpha\right)\right)\tilde{S}_{11}^{\mathrm{exp}}(\theta)
\end{equation}
This formal scattering matrix is very similar to the Bullough-Dodd
scattering matrix (\ref{eq:BD_S-matrix}) and satisfies the bootstrap
equation $S_{11}(\theta+i\frac{\pi}{3})S_{11}(\theta-i\frac{\pi}{3})=S_{11}(\theta)$.
Comparing, however, to the perturbative expansion (\ref{eq:pertS})
we can see a mismatch, thus we restrict our analysis only to the well-founded
form (\ref{eq:S11}). The $\gamma=\frac{\alpha}{2}$ choice would
lead to the following mass of the quantum excited kink: 
\begin{equation}
M^{*}=M\cos\alpha+m_{1}\cos\frac{\pi}{6}
\end{equation}

\begin{figure}[H]
\centering{}\includegraphics[height=4cm]{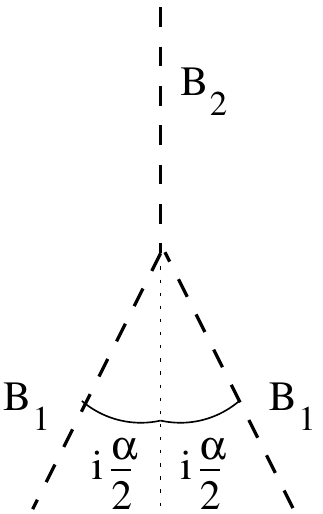}\hspace{2cm}\includegraphics[height=4cm]{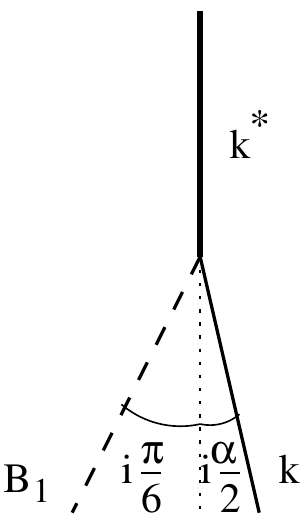}
\protect\caption{The $B_{1}-B_{1}$ fusion process, together with the $B_{1}$-kink
fusing to excited kink process.}
\label{kinkex}
\end{figure}

Using that $m_{2}=2m_{1}\cos\frac{\alpha}{2}=2M\sin\alpha$ , we can
see that $\beta_{2}=\pi-2\alpha$. From the fusion we also obtain
\begin{equation}
S_{k2}(\theta)=S\left(\theta,i\frac{\pi}{2}-i\alpha\right)S_{k2}^{\mathrm{exp}}(\theta)
\end{equation}

Our analysis is very natural from an integrable point of view, moreover,
a similar parametrization comes out from semiclassical investigations,
too \cite{Mussardo:2006iv}. In the following we use these masses
and scattering matrices to describe the finite size spectrum, which
we compare to the numerical data. This will decide whether such bootstrap
based proposals are adequate or not.

\subsection{Polynomial finite size corrections}

The leading finite size corrections are polynomial in the inverse
of the volume and are due to momentum quantization. If the volume
is large compared to the particle number then particles travel freely
most of the time, except when they scatter on each other such that
their wave functions pick up the scattering phases. Summing up all
these phases when they move around the circle and adding the translational
phase, the result should be an integer/half-integer multiple of $2\pi$,
ensuring that the wave function is periodic/antiperiodic: 
\begin{equation}
e^{ip_{j}L}\prod_{n:n\neq j}S_{jn}(\theta_{j}-\theta_{n})=\pm1
\end{equation}
This equation is called the Bethe-Yang equation and is valid for integrable
models, where particle numbers cannot change. We believe that in certain
kinematical domains it gives a good description of the spectrum even
for non-integrable models. Once the momenta are calculated, the energy
is simply 
\begin{equation}
E=\sum_{n=1}^{N}m_{n}\cosh\theta_{n}
\end{equation}
In the following, we elaborate this description for small particle
numbers.

For a one-particle state it implies that 
\begin{equation}
e^{im_{j}L\sinh\theta}=1\quad;\quad j=1,2
\end{equation}
Taking its logarithm, we can see how the momentum is quantized in
a volume $L$: 
\begin{equation}
m_{j}\sinh\theta=p_{j}=\frac{2\pi}{L}n
\end{equation}
This is identical to the free momentum quantization, thus the one-particle
levels get only exponentially small, interaction-dependent finite
size corrections. In the zero-momentum sector with periodic boundary
condition we have the standing $B_{1}$ and $B_{2}$ particles, i.e.
with $n=0$. For antiperiodic boundary condition we have standing
kink states. 

For a state with two $B_{1}$ particles having opposite momenta $p=\pm m_{1}\sinh\theta$,
the equation implies that 
\begin{equation}
e^{im_{1}L\sinh\theta}S_{11}(2\theta)=1\label{eq:2ptBY}
\end{equation}
Taking its logarithm we arrive at 
\begin{equation}
m_{1}L\sinh\theta-i\log S_{11}(2\theta)=2\pi n\quad;\qquad n\in\mathbb{Z}\label{eq:logBY}
\end{equation}
Since the $\theta\to-\theta$ change brings (\ref{eq:logBY}) to the
same equation but for $n\to-n$, we assume that $n\geq0$. Once this
equation is solved at a given $L$ and $n$ for $\theta$, denoted
by $\theta_{n}(L)$, the energy of the corresponding two-particle
state is 
\begin{equation}
E_{2}(n,L)=2m_{1}\cosh\left(\theta_{n}(L)\right)\label{eq:2ptBYenergy}
\end{equation}

In theories where there is a bound state pole in the scattering matrix,
the equation (\ref{eq:logBY}) admits a complex solution. Indeed,
let us assume that in eq. (\ref{eq:2ptBY}) the rapidity $\theta$
can be written as $\theta=iu$ with $u$ being real. Clearly the factor
$e^{-m_{1}L\sin u}$ goes to zero in the large volume limit. As the
product is $1$, the S-matrix contributions should diverge, which
forces $2u=\alpha+\delta u(L)$, where $\delta u(L)$ goes to zero
in the large $L$ limit. We can even calculate how it vanishes by
expanding eq. (\ref{eq:2ptBY}):
\begin{equation}
e^{-m_{1}L\sin\frac{\alpha}{2}}\frac{\Gamma^{2}}{2\delta u}=1\quad\to\qquad\delta u=\tan(\alpha)e^{-m_{1}L\sin\frac{\alpha}{2}}
\end{equation}
The energy of this state is 
\begin{equation}
E_{2}(0,L)=2m_{1}\cos\frac{\alpha}{2}-m_{1}\sin\frac{\alpha}{2}\tan(\alpha)e^{-m_{1}L\sin\frac{\alpha}{2}}
\end{equation}
Clearly it describes a standing $B_{2}$ particle together with its
leading exponential finite size corrections (\ref{eq:muterm}). The
equation (\ref{eq:logBY}) sums up all of these exponentially small
corrections for $n=0$. 

Similar analysis can be performed in the antiperiodic sector using
the Bethe-Yang equations of the kink-breather scatterings. There the
breathers are quantized as 
\begin{equation}
m_{1}L\sinh\theta-i\log S_{1k}(\theta_{1}-\theta_{k})=\pi(2n+1)\quad;\qquad n\in\mathbb{Z}\label{eq:logBYap}
\end{equation}

\subsection{Leading exponential corrections}

The leading finite size correction for the vacuum is described by
\begin{equation}
E_{0}(L)=-m\int_{-\infty}^{\infty}\frac{d\theta}{2\pi}\cosh\theta\, e^{-mL\cosh\theta}\label{eq:vacLuscher}
\end{equation}
where $m$ is the mass of the lightest particle in the spectrum. This
correction is related to the virtual process in which a particle--antiparticle
pair appears from the vacuum, travel around the world and annihilate
when they meet again. 

The general form of the leading mass corrections was calculated by
Lüscher \cite{Luscher:1985dn}. It has two terms: The $F$-term corresponds
to the process in which a virtual particle--antiparticle pair of type
$b$ appears from the vacuum and after traveling around the world
and scattering on particle type $a$, they annihilate. The $\left(-1\right)$
is the subtraction of a similar process for the vacuum. 
\begin{equation}
\Delta_{F}m_{a}=-m_{b}\int_{-\infty}^{\infty}\frac{d\theta}{2\pi}\cosh\theta\,\left(S_{ba}\left(i\pi/2+\theta\right)\right)-1)e^{-m_{b}L\cosh\theta}\label{eq:Fterm}
\end{equation}
The $\mu$-term is the residue of the $F$-term at the bound state
pole: 
\begin{equation}
\Delta_{\mu}m_{a}=-\sum_{b,c}\theta\left(m_{a}^{2}-\vert m_{b}^{2}-m_{c}^{2}\vert\right)\mu_{ab}^{c}(-i)\mbox{Res}S_{ab}(\theta)e^{-\mu_{ab}^{c}L}\label{eq:muterm}
\end{equation}
where $\mu_{ab}^{c}$ is the altitude of the mass triangle with base
$m_{c}$, and it is being understood that $c$ appears as a bound
state of $a$ and $b$. Physically it corresponds to the process in
which a particle of type $a$ decays virtually into the particles
$b$ and $c$, which both travel forward in time. After traveling
around the finite world they meet again on the other side and fuse
back to particle $a$.

\section{Numerical analysis and interpretations}

In this section we summarize our numerical results. The paper \cite{Coser:2014lla}
used the conformal normal ordering scheme. To test our numerical implementations,
we first developed a program in this convention and successfully reproduced
their weak coupling data. Then, to avoid infrared divergences for
large volumes, we switched to the $(m,L)$ normal ordering conventions
(\ref{eq:g2g4}). All of the data in this section are in this $(m,L)$
convention. We plot the dimensionless energies $E/m$ against the
dimensionless volume $l=mL$ and we analyze both the periodic and
the antiperiodic sector.

\subsection{Periodic boundary condition}

\subsubsection{Breather sector}

\begin{figure}[H]
\begin{centering}
\includegraphics[width=10cm]{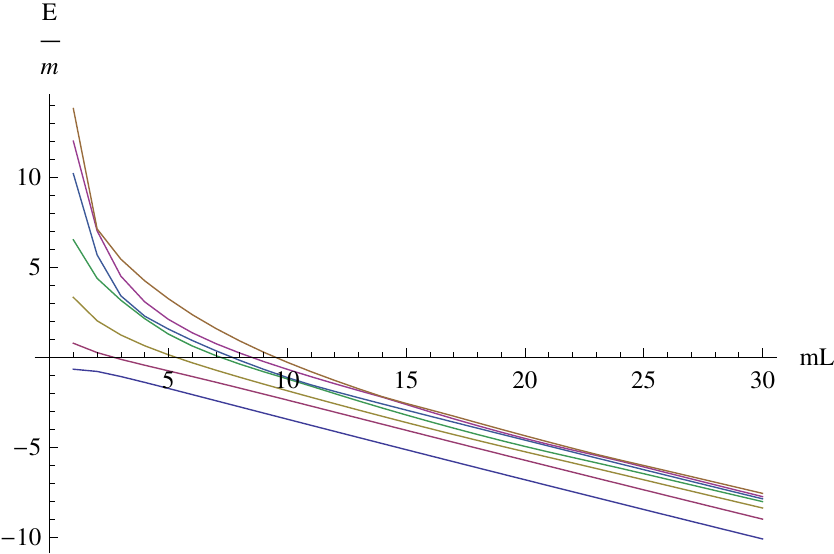}
\par\end{centering}

\protect\caption{The raw energy spectrum of the model as the function of the volume }

\centering{}\label{rawspec}
\end{figure}

We first show a typical spectrum at $g=1.2$ on Figure \ref{rawspec}.

Clearly the system has a ground state energy density such that for
large volumes all energy levels contain a term $E_{bulk}L$. This
term can be analyzed already on the ground state, which goes as $E_{0}(L)=E_{\mathrm{bulk}}L+E_{exp}(L)$.
Here $E_{exp}(L)$ is exponentially small for large $L$. We normalized
the energy density to zero for the free boson with mass $m$. But
once the interaction is switched on, the ground state energy density
becomes a measurable physical quantity.

\begin{figure}[H]
\begin{centering}
\includegraphics[width=10cm]{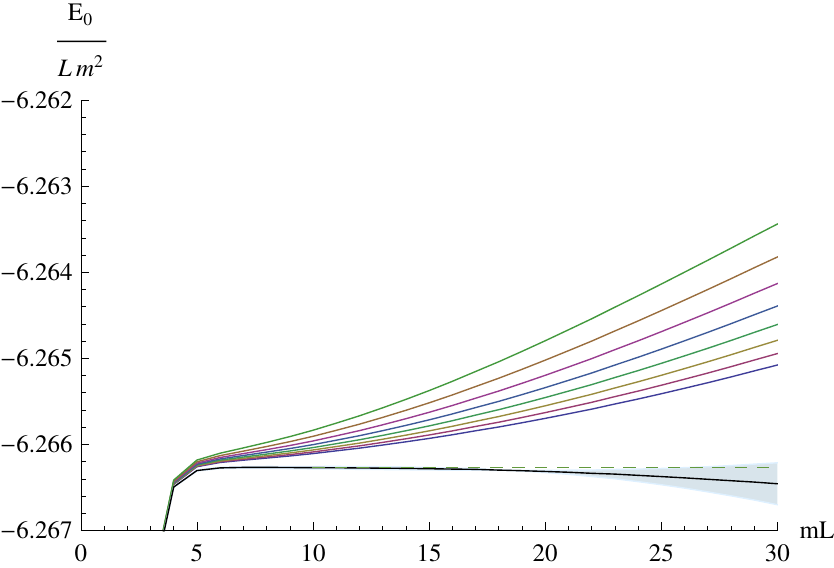}
\par\end{centering}

\protect\caption{The ground-state energy density $e_{0}(l)/l$ as the function of the
volume for various truncated energy levels together with the extrapolated
results. The energy density obtained at $l=7$ is indicated with a
dashed line.}

\centering{}\label{figcutoffdep}
\end{figure}

In measuring every quantity we should keep in mind that we determine
the spectrum at various truncation levels $e_{cut}$. The dependence
of the energy density on the cut-off is demonstrated on Figure \ref{figcutoffdep}. 

As truncation effects are very large, we have to extrapolate in the
cut-off dependence in order to get a reasonable result. Based on the
analysis in \cite{Rychkov:2014eea} we found it useful to fit the
following function 
\begin{equation}
E(e_{cut})=a+b\frac{\log^{2}e_{cut}}{e_{cut}^{2}}+c\frac{\log e_{cut}}{e_{cut}^{2}}\label{eq:ecutlog2}
\end{equation}
In order to say something about the accuracy of the extrapolation,
we also fitted a simpler form

\begin{equation}
E(e_{cut})=a+b\frac{\log^{2}e_{cut}}{e_{cut}^{2}}\label{eq:ecutlog2-1}
\end{equation}
and used the difference of the two results for the error approximation
on Figure \ref{figcutoffdep}.

\begin{figure}[H]
\begin{centering}
\includegraphics[width=10cm]{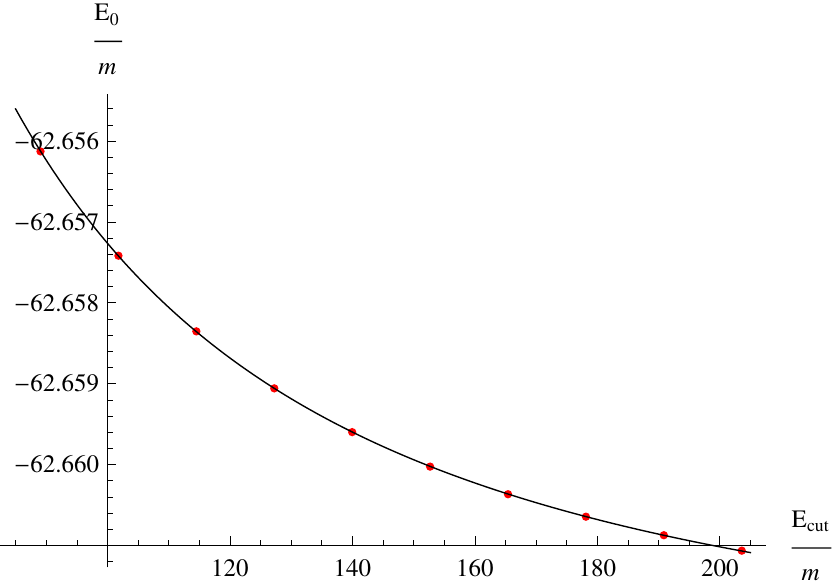}
\par\end{centering}

\protect\caption{Data points as a function of the truncation energy together with the
fitted function for the ground-state energy density at $g=0.06$ and
volume $l=10$. }
\label{log2ecut}
\end{figure}

We show on Figure \ref{log2ecut} the convergence of the energy density
values along with the fitted function. 

\begin{figure}[H]
\begin{centering}
\includegraphics[width=10cm]{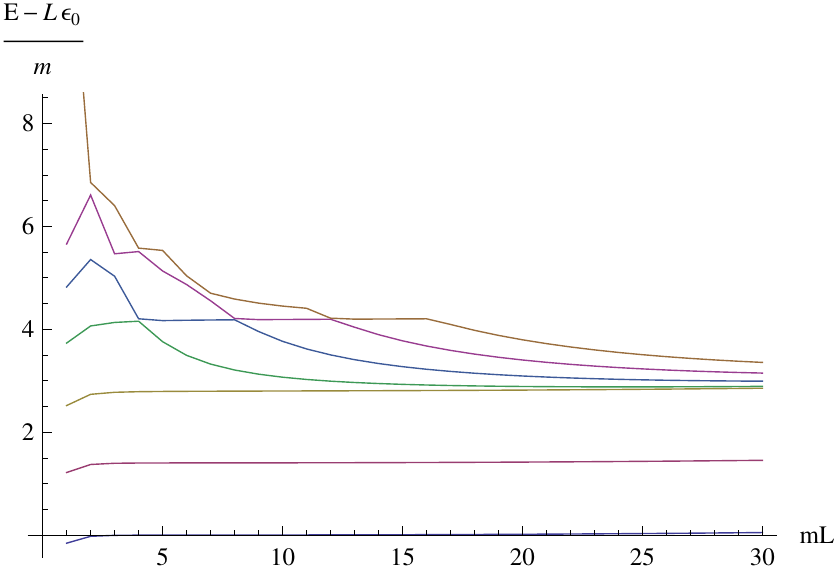}
\par\end{centering}

\protect\caption{Low lying spectrum at $g=0.06$: the first seven energy levels (with
the vacuum energy density subtracted) are plotted against the volume
in the periodic sector. The splitting of even and odd lines is so
small that it is not visible at this scale.}

\label{typspec}
\end{figure}

In the following we use (\ref{eq:ecutlog2}) to extrapolate each energy
level and show on the plots the extraploted value only. In most of
the cases we subtract the vacuum energy density, such that a typical
spectrum looks like the one on Figure \ref{typspec}. 

We can easily identify the first and second breathers and even a resonance
around $3m_{1}$.

\begin{figure}[H]
\begin{centering}
\includegraphics[width=10cm]{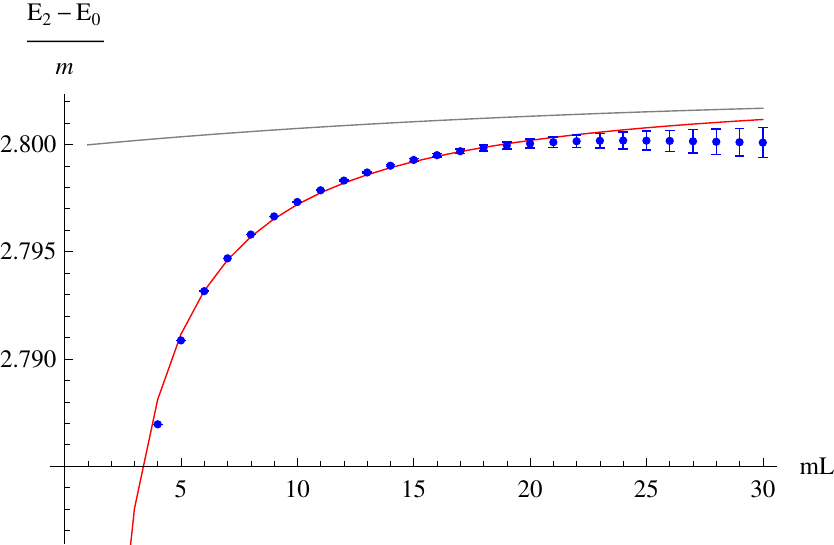}
\par\end{centering}

\protect\caption{The imaginary Bethe-Yang description of $B_{2}$ at $g=0.06$, together
with the leading finite size correction (shown in gray).}

\centering{}\label{B2}
\end{figure}

Let us start with the standing particles. For the $B_{1}$ particle
the energy is a constant function of the volume up to exponential
finite size corrections, however for $B_{2}$ we observed stronger
finite size corrections. We blow up this part of the spectrum on Figure
\ref{B2}. The reason is that $B_{2}$ is a weakly coupled bound state
of two $B_{1}$ particles. In decreasing the volume we pump energy
into the system, which makes the bound state dissolve and behave more
like two $B_{1}$ states. That is why the equation (\ref{eq:logBY})
with imaginary rapidities, shown by a red solid line, provides a very
precise description. This fitting is very sensitive to $\alpha$,
which turns out to be $\alpha=0.06$.

\begin{figure}[H]
\begin{centering}
\includegraphics[width=10cm]{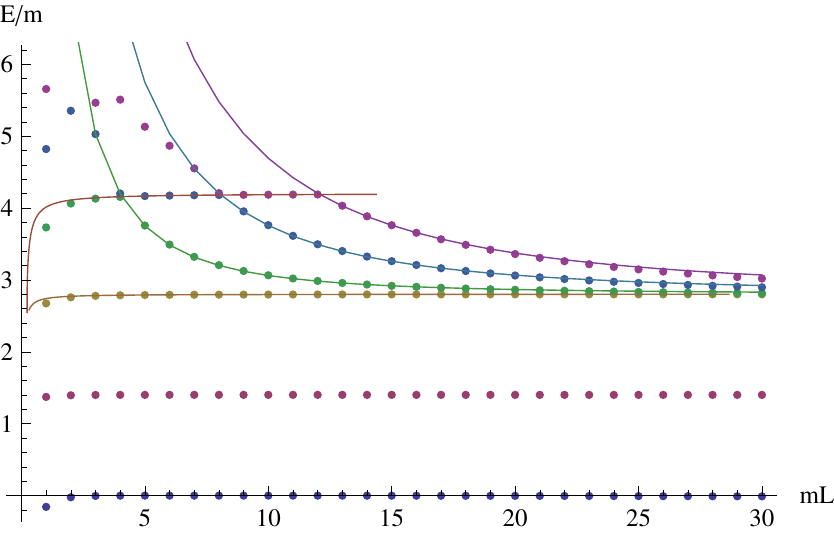}
\par\end{centering}

\protect\caption{Low lying even spectrum at $g=0.06$: the first six energy levels
(with the vacuum energy density subtracted) are plotted against the
volume together with the solid Bethe-Yang lines for $n=0,1,2,3$.
The resonance is well described by a three-particle imaginary BY line
including a stationary breather.}

\centering{}\label{BYspectrum}
\end{figure}

The other lines in the spectrum are described by the Bethe-Yang equations
with real rapidities and quantization numbers $n=1,2,3$. These curves
are shown on Figure \ref{BYspectrum}. 

\begin{figure}[H]
\begin{centering}
\includegraphics[width=10cm]{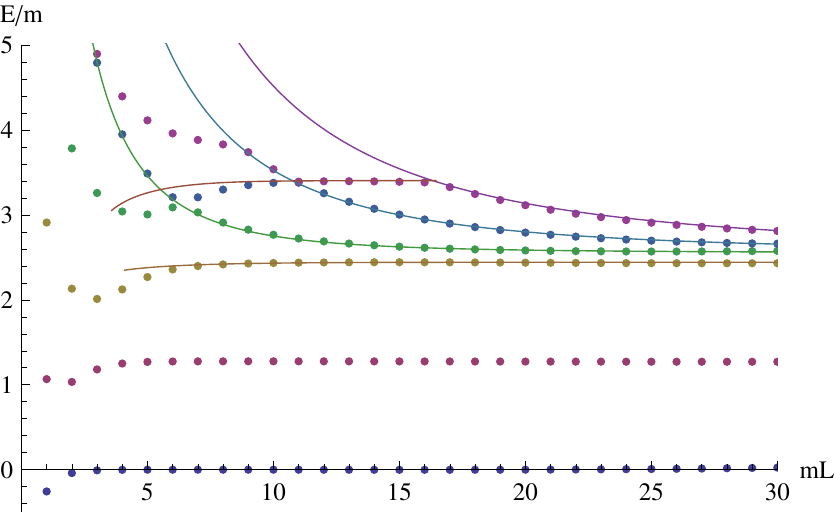}
\par\end{centering}

\protect\caption{Low lying spectrum at $g=0.6$: the first six even energy levels (with
the vacuum energy density subtracted) are plotted against the volume
together with the solid Bethe-Yang lines for $n=0,1,2,3$ and the
three-particle BY line for the resonance. }

\centering{}\label{g06spec}
\end{figure}

The spectrum for small $g$ can be very precisely described in a similar
manner in terms of $m_{1}$ and $\alpha$. On Figure \ref{g06spec}
we show a similar plot at $g=0.6$. 

\begin{figure}[H]
\begin{centering}
\includegraphics[width=10cm]{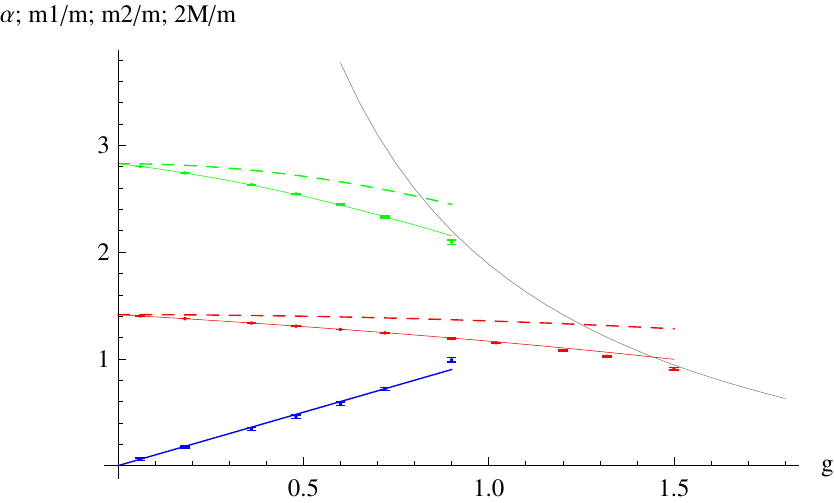}
\par\end{centering}

\protect\caption{The masses of the breathers, $m_{1}$ (red), $m_{2}$ (green), together
with the scattering parameter $\alpha$, (blue) as functions of the
coupling. The numerical $B_{2}$ mass is computed using eq. (\ref{eq:b2mass}).
The semiclassical results (\ref{eq:semimn},\ref{eq:semialpha}) are
indicated by dashed lines. The semiclassical result for $2M$ is shown
in grey. The perturbative correction to $m_{1}$ (\ref{eq:m1pert})
is indicated by a solid red line. The solid green line depicts $m_{2}$
as calculated by (\ref{eq:b2mass}), using the perturbative result
for $m_{1}$ and the semiclassical value of $\alpha$.}

\centering{}\label{m12agdep}
\end{figure}

Performing a similar analysis we demonstrated for $g=0.06$ we can
extract the coupling dependence of $m_{1}$, $m_{2}$ and $\alpha$.
This is shown on Figure \ref{m12agdep}. 

We can compare the mass of the elementary excitation to perturbation
theory \cite{Rychkov:2015vap}:
\begin{equation}
m_{pert}^{2}(g)-m_{0}^{2}=m_{0}^{2}\left(-\frac{g}{2\sqrt{3}}-\frac{1}{36}\left(0.375+\left(2.2492+2.8020\right)-\frac{1}{2}\left(1.06864+1.9998+5.50025\right)\right)g^{2}\right)\label{eq:m1pert}
\end{equation}

Observe that the perturbative formula starts at linear order in $g$
and gives a very good approximation, opposed to the semiclassical
formula, which starts at quadratic order. Having a closer look at
this quantum spectrum, we can see the following: For small couplings,
we have two breathers and a very heavy pair of kinks. As the coupling
increases, the measured masses are smaller than the semiclassical
values and the second breather disappears from the spectrum when it
can decay into a kink-antikink pair, i.e. when $m_{2}(g)=2M(g)$,
this happens around $g=0.9$. Similarly, the light breather, the elementary
excitation of the quantum field disappears at around $g=1.5$, where
its mass equals twice the mass of the kink: $m_{1}(g)=2M(g)$. For
small couplings, the semiclassical calculations are reliable, which
exclude the existence of more neutral particles. Indeed we can only
see a resonance around $3m_{1}$. This confirms the conjecture of
\cite{Mussardo:2006iv}. 

It is interesting to take a look at the spectrum at stronger couplings,
when the breathers disappear from the spectrum.

\begin{figure}[H]
\begin{centering}
\includegraphics[width=10cm]{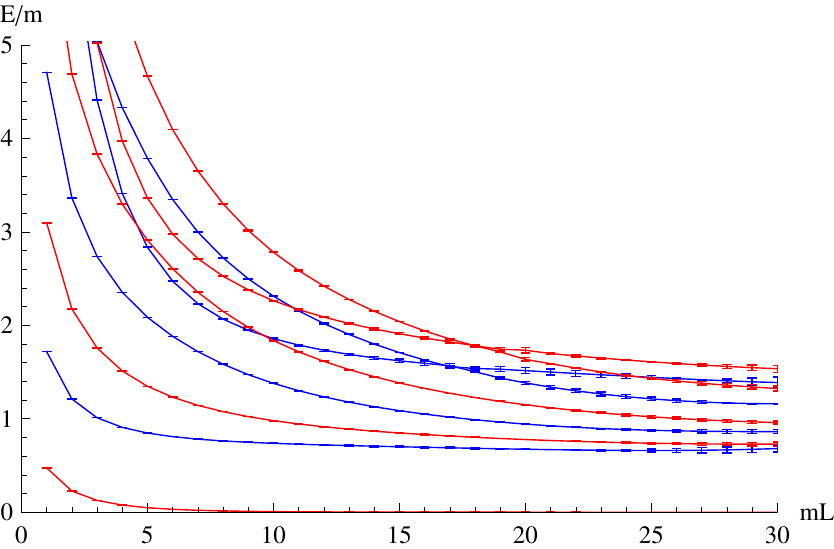}
\par\end{centering}

\protect\caption{Low lying spectrum at $g=1.8$: The low lying even (blue) and odd
(red) energy levels (with the vacuum energy density subtracted) are
plotted against the volume.}

\centering{}\label{specg1.8}
\end{figure}

\begin{figure}[H]
\begin{centering}
\includegraphics[width=7cm]{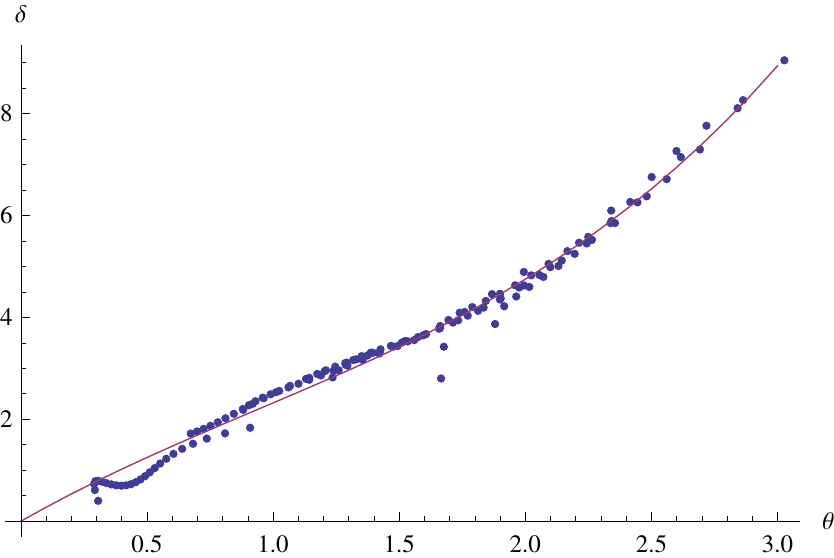}~~~~~~~\includegraphics[width=7cm]{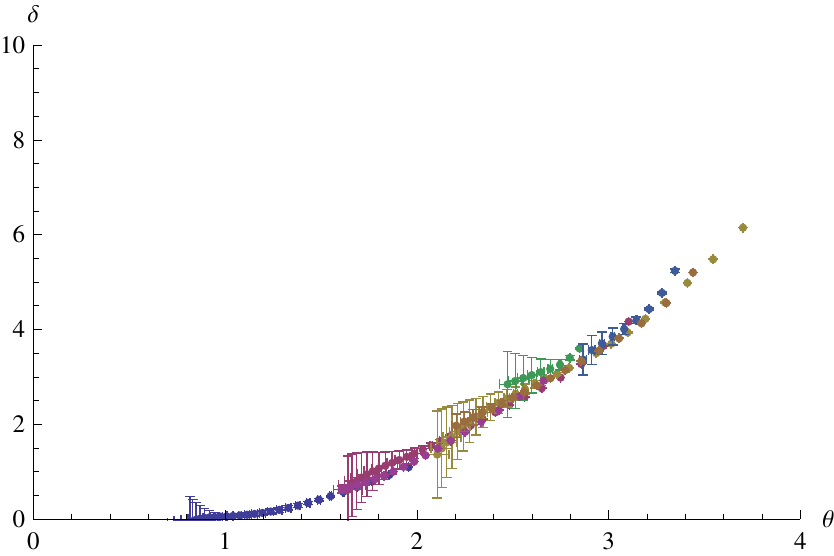}
\par\end{centering}

\protect\caption{Kink-anti-kink scattering phase shifts as functions of the rapidity
for $g=1.8$ (left) and for $g=2.4$ (right). The various points were
extracted from lines with different $n$ in (\ref{eq:logBY}). For
$g=1.8$ we also presented a fitted third order polynomial in $\theta$
, while for $g=2.4$ the errors of our data. }

\centering{}\label{phaseshift}
\end{figure}

Such a spectrum is shown on Figure \ref{specg1.8} for $g=1.8$. Clearly
we can see two and four particle scattering states. By identifying
the points which correspond to the two particle Bethe-Yang lines with
quantization numbers $n$ and using (\ref{eq:2ptBY},\ref{eq:2ptBYenergy})
we can measure the $S_{k\bar{k}}$ scattering phase shift $\delta(\theta)$
as follows: We first pick up a quantization number $n$ and analyze
the $E_{2}(n,L)$ curve. For a given $L$ we transform the energy
using (\ref{eq:2ptBYenergy}) to rapidity
\begin{equation}
E_{2}(n,L)=2M\cosh\left(\theta_{n}(L)\right)\longrightarrow\theta_{n}(L)=\cosh^{-1}\left(\frac{E_{2}(n,L)}{2M}\right)
\end{equation}
We then use (\ref{eq:2ptBY}) to transform the volume for the given
$\theta$ to the phase shift as: 
\begin{equation}
-i\log S_{k\bar{k}}(2\theta)=2\pi n-ML\sinh\theta
\end{equation}
Parameterizing the kink-anti-kink scattering matrix as 
\begin{equation}
S_{k\bar{k}}(\theta)=-e^{i\delta(\theta)}
\end{equation}
the resulting phase-shifts are displayed for $g=1.8$ and $g=2.4$
in Figure \ref{phaseshift}. 

Observe that the phase shift is smaller and smaller once we increase
the coupling and approach the critical point. We can fit a third order
polynomial for the phase shift at $g=1.8$ as follows:
\begin{equation}
\delta(2\theta)=2.82(3)\,\theta-0.77(10)\,\theta^{2}+0.27(9)\,\theta^{3}
\end{equation}
 Clearly in the method above we needed the precise value of the kink
mass. We measured this mass with two different methods: first from
the splitting of the ground-state energy in subsection \ref{sub:vacuumsplitting},
then from the ground-state of the anti-periodic sector in section
\ref{sub:antiperiodic}. Both measurements indicated that the kink
mass is close to its semiclassical value in the whole range below
the critical point, but the errors become large at stronger couplings.
Therefore we used the semiclassical kink mass for these measurements.
Since we do not consider the error of the kink mass, the errors are
somewhat underestimated.

\begin{figure}[H]
\begin{centering}
\includegraphics[width=10cm]{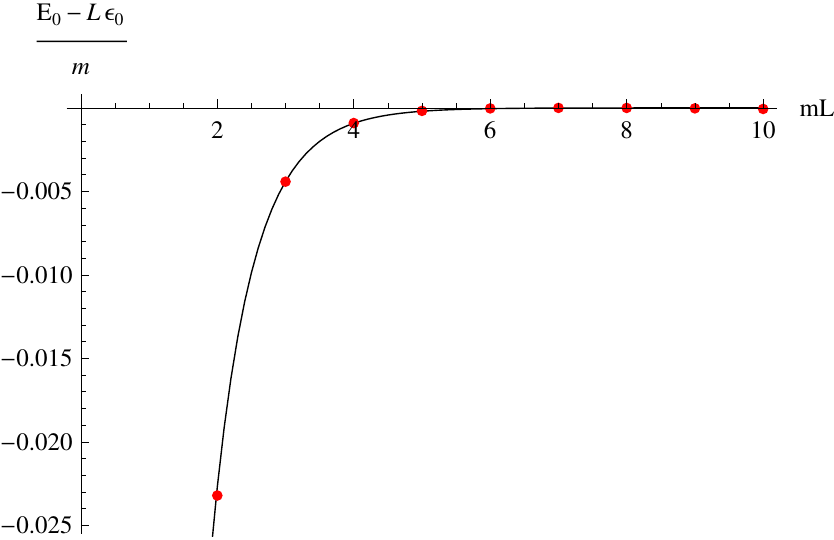}
\par\end{centering}

\protect\caption{Leading exponential correction of the vacuum energy at $g=0.06$. }

\centering{}\label{vacexp}
\end{figure}

So far we checked only the polynomial finite size corrections. Let
us investigate now the leading exponential corrections. For $g=0.06$,
the mass of the lightest particle is $m_{1}=1.40182(10)$, which can
be used to check the leading exponential correction of the vacuum
energy (\ref{eq:vacLuscher}). The comparision is shown on Figure
\ref{vacexp}, providing a convincing evidence. 

The similar comparision between the energy level of the first breather
and (\ref{eq:Fterm}) indicates that we are missing some interesting
physics. 

\begin{figure}[H]
\begin{centering}
\includegraphics{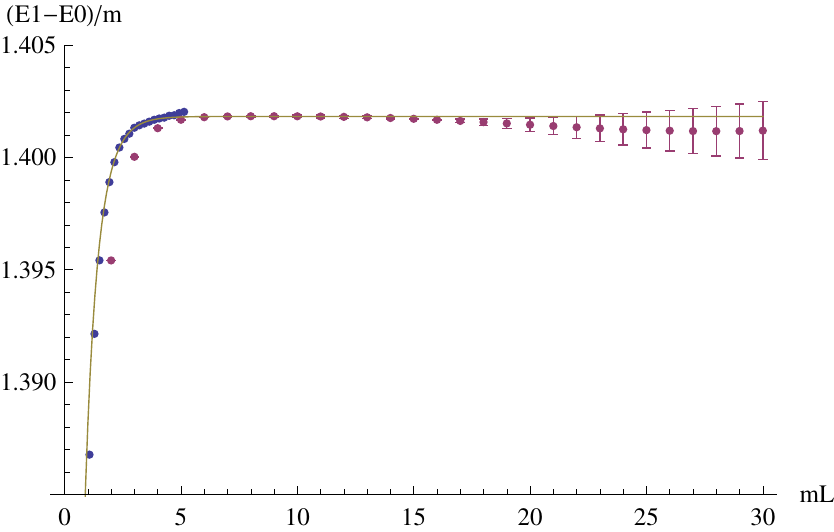}
\par\end{centering}

\protect\caption{The finite volume mass of the first breather together with the leading
finite size correction (\ref{eq:Fterm}) at $g=0.06$. $B_{1}$ energies
from the rescaled TCSA spectrum of the sine-Gordon theory at $p\pi=0.06$
are also indicated by blue dots. The latter fits very well to the
leading correction. }

\centering{}\label{B1Fterm}
\end{figure}

\begin{figure}[H]
\begin{centering}
\includegraphics{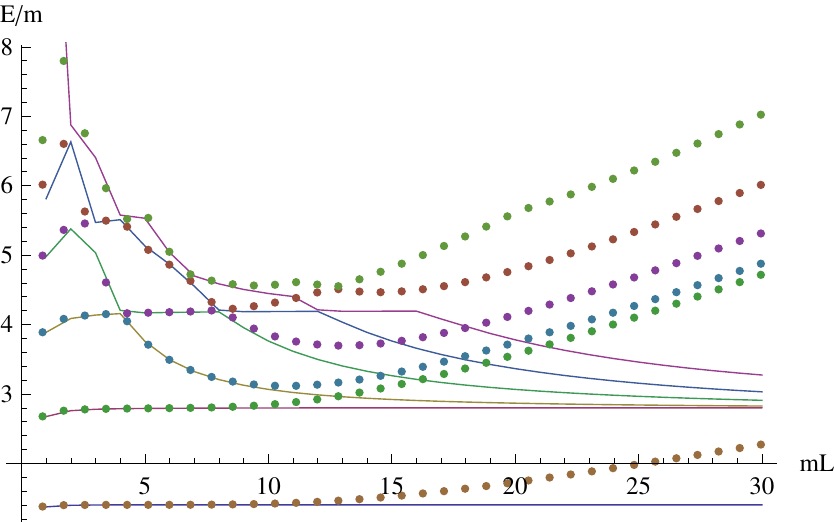}
\par\end{centering}

\protect\caption{The rescaled TCSA spectrum of the sine-Gordon theory at $p\pi=0.06$
and $e_{cut}=20$, together with the spectrum of the $\phi^{4}$ theory
at $g=0.06$. The $\phi^{4}$ data are indicated by solid lines. All
data are normalized wrt. the corresponding ground state energies.}

\centering{}\label{sGvsphi4}
\end{figure}

One may wonder if summing up all higher corrections coming from the
elastic scattering matrix $S_{11}(\theta)=S(\theta,i\alpha)$ together
with the contributions of the second breather can save the situation.
To test this idea, we can compare the spectrum of our model to the
spectrum of the sine-Gordon theory, at the coupling where the first
breather's scattering matrix agrees with $S_{11}(\theta)$. We used
the TCSA program from \cite{Bajnok:2000wm} and rescaled the energies
and volume such that the dimensionless quantities agree with this
paper's convenction. (In the TCSA program the soliton mass is normalized
to $1$.) The two spectra are plotted on the same figure on Figure
\ref{sGvsphi4}. Clearly the spectra agree very precisely for smaller
volumes, when the error of the TCSA analysis is small. (We did not
perform any extrapolation in the TCSA energy cut which was chosen
to be $20$.) One surprising outcome is that the leading Lüscher correction
(\ref{eq:Fterm}) agrees very well with the TCSA data, but not with
the $\phi^{4}$ ones. It indicates that the inelastic processes play
an important role in determining the finite size effects. 

\subsubsection{Kink mass from the vacuum splitting \label{sub:vacuumsplitting}}

So far we analyzed the even energy levels only. To gain information
about the kink mass, one can analyze the gap between the even and
odd ground-state energies \cite{Bajnok:2000wm,Rychkov:2015vap}: 
\begin{equation}
E_{0}^{even}(L)-E_{0}^{odd}(L)=\sqrt{\frac{2M}{\pi L}}e^{-ML}+\dots
\end{equation}

\begin{figure}[H]
\begin{centering}
\includegraphics[width=10cm]{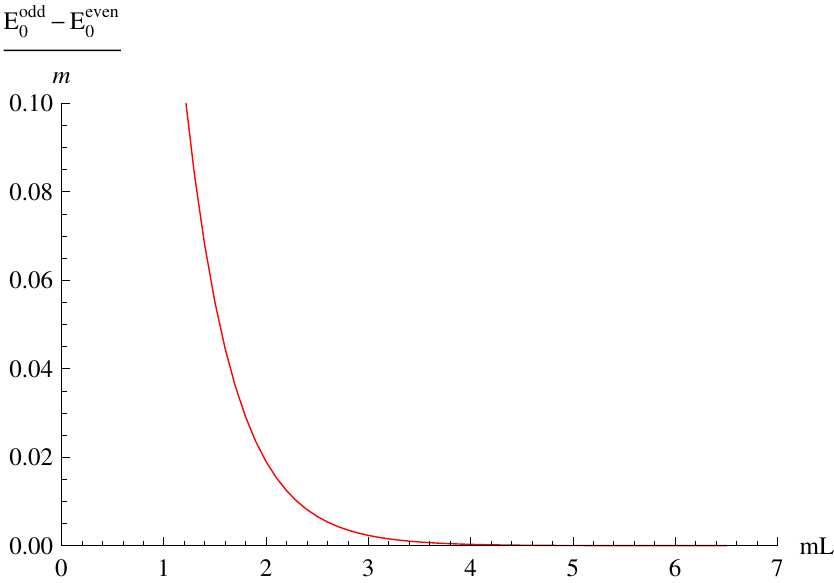}
\par\end{centering}

\protect\caption{Difference of the ground-state energies at $g=0.6$ of the even and
the odd sector. }

\centering{}\label{evenoddvac}
\end{figure}

\textcolor{black}{The splitting is demonstrated on Figure \ref{evenoddvac}. }

\begin{figure}[H]
\begin{centering}
\includegraphics[width=10cm]{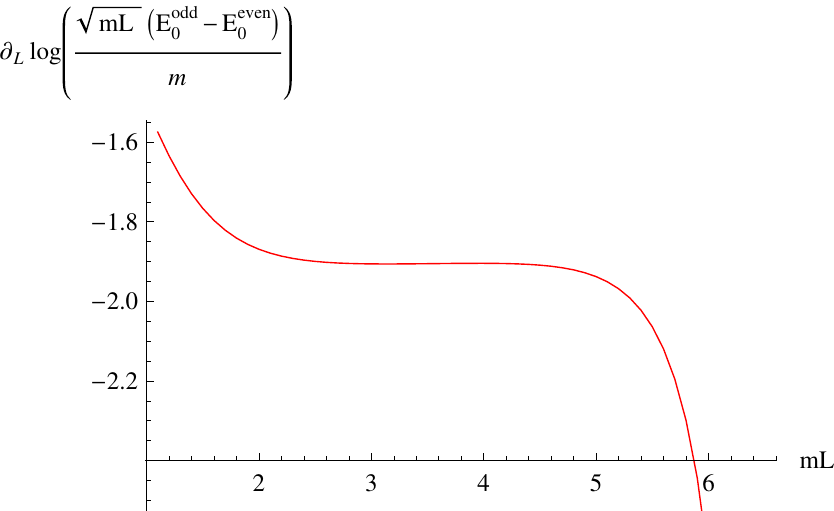}
\par\end{centering}

\protect\caption{The slope of the logarithm of the ($\sqrt{L}$-multiplied) splitting
between the even and odd ground-state energies. ($g=0.6)$ }

\centering{}\label{logslope}
\end{figure}

In order to extract the kink mass, we first multiply the numerical
data by $\sqrt{L}$, then we measure the slope of the logarithm of
the resulting curve by numerical differentiation. The result is shown
on Figure \ref{logslope}. 

\begin{figure}[H]
\begin{centering}
\includegraphics[width=10cm]{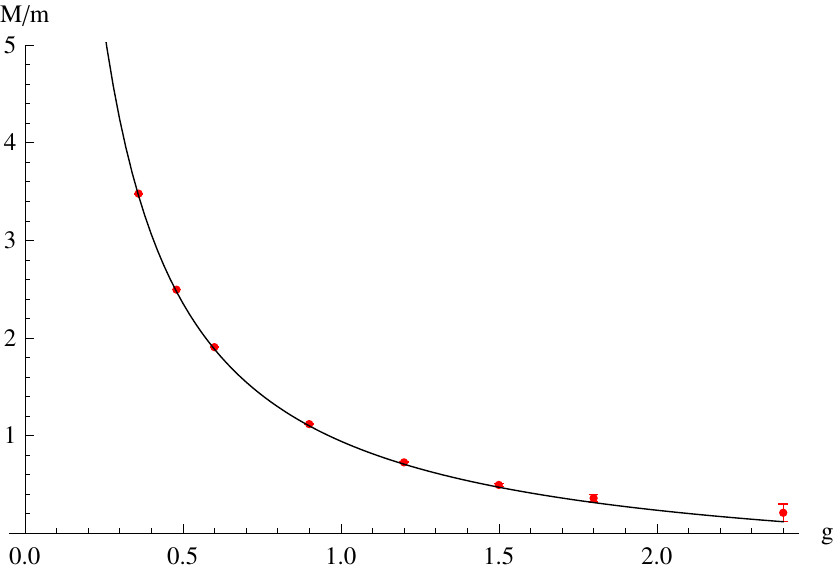}
\par\end{centering}

\protect\caption{Kink mass as a function of the coupling extracted from the splitting
together with the semiclassical result. }

\centering{}\label{kinkmassgap}
\end{figure}

Using this analysis, we can obtain the kink mass as a function of
the coupling. The curve on figure \ref{logslope} seems to be monotonic.
However, it actually has a little local minimum and a local maximum
which can be used to estimate the error. The measured kink masses
for different couplings are shown on Figure \ref{kinkmassgap}. 

\subsection{Antiperiodic boundary condition\label{sub:antiperiodic}}

\begin{figure}[H]
\begin{centering}
\includegraphics[width=10cm]{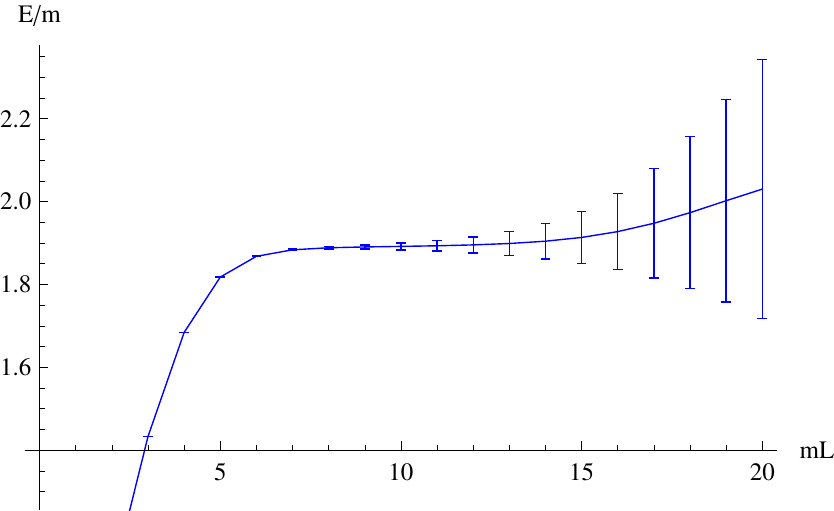}
\par\end{centering}

\protect\caption{The lowest lying state in the antiperiodic sector at $g=0.6$.}

\centering{}\label{kinkmass}
\end{figure}

The lowest lying state in the antiperiodic sector is the kink state.
After subtracting the periodic vacuum, we found the result presented
on Figure \ref{kinkmass}. 

\begin{figure}[H]
\begin{centering}
\includegraphics[width=10cm]{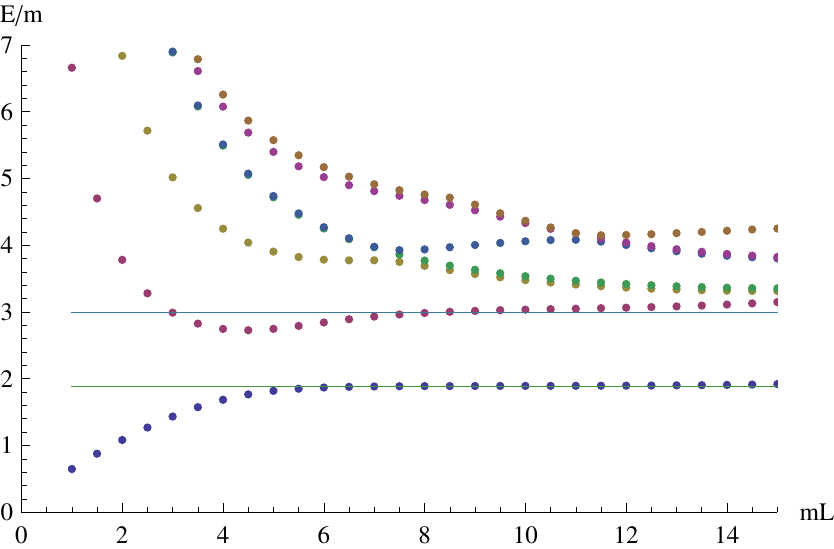}
\par\end{centering}

\protect\caption{The low lying spectrum in the antiperiodic sector at $g=0.6$. We
indicated the infinite volume masses of the semiclassical kink and
excited kink by horizontal green and blue lines, respectively.}

\label{lowkinksector}
\end{figure}

The low lying spectrum in the kink-sector is presented on Figure \ref{lowkinksector}.

\begin{figure}[H]
\begin{centering}
\includegraphics[width=10cm]{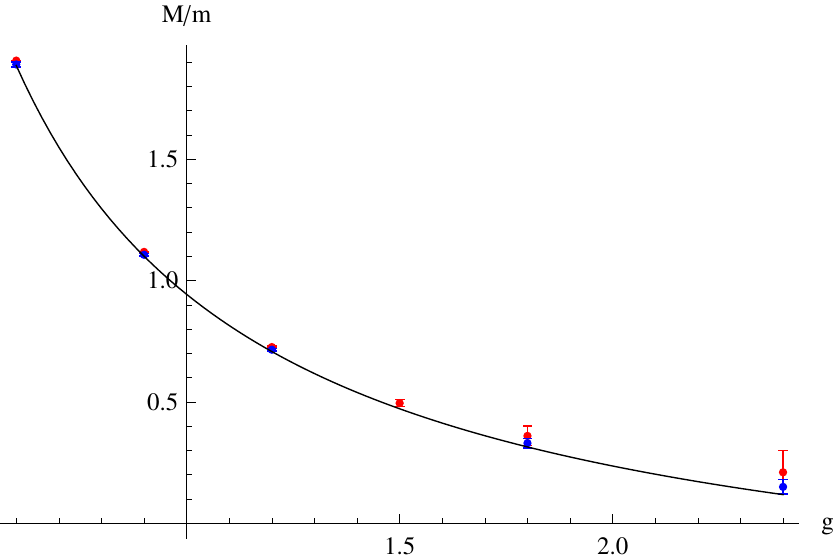}
\par\end{centering}

\protect\caption{Kink mass as a function of the coupling extracted in the antiperiodic
sector (blue) together with the results from the periodic sector (red)
and from semiclassical considerations (solid line). }

\centering{}\label{kinkmassall}
\end{figure}

 From the antiperiodic sector's ground state we can extract the kink
mass. It is in complete agreement with the one we obtained from the
ground state splitting and both are very close to the semiclassical
result. These data are shown on Figure \ref{kinkmassall}. 

\subsection{Critical point and the spectrum of the Ising model}

\begin{figure}
\centering{}\includegraphics{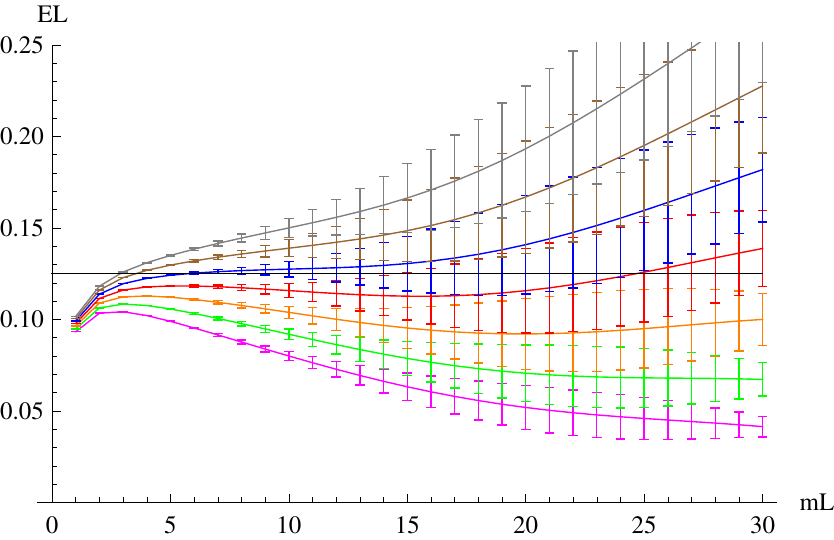}\protect\caption{Relative energies of the lowest state in the odd sector, multiplied
by $l/2\pi$ for couplings g=2.76 (purple), g=2.88 (green), g=3 (orange),
g=3.12 (red), g=3.24 (blue), g=3.36 (brown) and g=3.48 (gray). The
line $el/(2\pi)=1/8$ is shown as a horizontal line.\label{fig:critloc}}
\end{figure}

The vanishing of the kink mass indicates that there is a second order
phase transition around $g=3$. However, increasing the coupling above
$g=1.8$ renders both direct kink mass measurement methods inaccurate.
On the other hand, it is known that at the critical point the spectrum
must be governed by a conformal field theory, which has the spectrum
\[
H_{CFT}=\frac{2\pi}{L}(L_{0}+\bar{L}_{0}-\frac{c}{12})\quad;\quad P_{CFT}=\frac{2\pi}{L}(L_{0}-\bar{L}_{0})
\]
Here $L_{0}$ and $\bar{L}_{0}$ are the Virasoro generators, which
measure the left and right conformal dimensions, respectively. Their
sum is the scaling dimension, while their difference is the conformal
spin, which is proportional to the momentum. As we analyze the zero
momentum sector of the theory we list the low lying scaling dimenions
of the periodic Ising model in this sector in Table \ref{Isingspectrum}.
\begin{table}
\begin{centering}
\begin{tabular}{|c|c||c|c||c|c|}
\hline 
state & scaling dim. & state & scaling dim. & state & scaling dim.\tabularnewline
\hline 
\hline 
$\vert0\rangle$ & $0$ & $\vert\frac{1}{2},\frac{1}{2}\rangle\phantom{\frac{L_{L}^{L}}{L_{L}^{L}}}$ & $1$ & $\vert\frac{1}{16},\frac{1}{16}\rangle$ & $\frac{1}{8}$\tabularnewline
\hline 
$L_{-2}\bar{L}_{-2}\vert0\rangle$ & $4$ & $L_{-1}\bar{L}_{-1}\vert\frac{1}{2},\frac{1}{2}\rangle\phantom{\frac{L_{L}^{L}}{L_{L}^{L}}}$ & $3$ & $L_{-1}\bar{L}_{-1}\vert\frac{1}{16},\frac{1}{16}\rangle$ & $\frac{1}{8}+2$\tabularnewline
\hline 
\end{tabular}
\par\end{centering}

\protect\caption{Low lying energy levels in the conformal Ising model in the zero momentum
sector with periodic boundary conditions.}

\label{Isingspectrum}
\end{table}
 We compare this spectrum with $\frac{l}{2\pi}$ times our spectrum.
First we plot the relative energies of the lowest state in the odd
sector against the volume for different couplings between $g=2.76$
and $g=3.48$. This line should correspond to the $\vert\frac{1}{16},\frac{1}{16}\rangle$
state at the point of the phase transition. The result is shown on
Figure \ref{fig:critloc}. This indicates that the critical point
is between $g=3.1$ and $g=3.3$.

\begin{figure}[H]
\begin{centering}
\includegraphics[width=10cm]{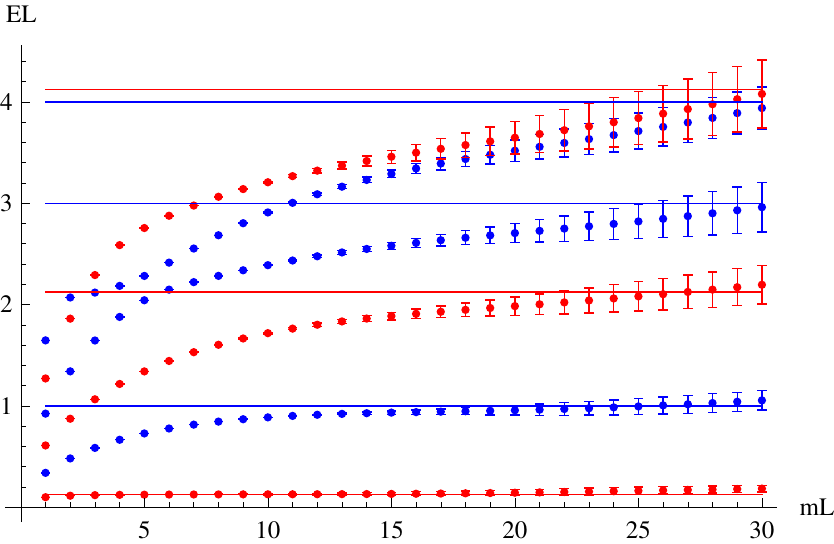}
\par\end{centering}

\protect\caption{Scaling dimensions of the periodic Ising model in the zero momentum
sector (solid lines), together with our spectrum in units of $\frac{2\pi}{l}$.
Blue lines and points are from the even, while red lines and points
are from the odd sector. }

\centering{}\label{Isingper}
\end{figure}

We plotted the low-lying periodic spectrum at $g=3.24$ in Figure
\ref{Isingper}. Similarly to the critical behaviour of the two-frequency
sine-Gordon model \cite{Bajnok:2000ar}, we nicely recover the conformal
Ising spectrum. Higher energy states approach their asymptotic values
later, but all the multiplicities are correct. This result is a very
strong confirmation that the critical behaviour is controlled by the
conformal Ising model. 

\subsection{Comparison to \cite{Rychkov:2015vap}}

In this subsection we compare our methods and results to those in
\cite{Rychkov:2015vap}. We focus on the periodic sector only as \cite{Rychkov:2015vap}
analyzed this sector. In order to facilitate the comparison we recall
that the quartic coupling in \cite{Rychkov:2015vap} is denoted by
$g_{4}\equiv g_{[14]}$, while the quadratic coupling is $\frac{1}{4}M_{[14]}^{2}$.
Since both papers use the same scheme, it implies that $M_{[14]}=m_{0}=\sqrt{2}m$
and $g=12\frac{g_{[14]}}{M_{[14]}^{2}}$. Consequently the dimensionless
volumes are related as $l_{[14]}=M_{[14]}L=\sqrt{2}l$, while the
energies as $\frac{H}{m}=\sqrt{2}\frac{H_{[14]}}{M_{[14]}}$

In the periodical case both papers used the same mini-Hilbert space
approach, but the numerical implementations of improving the cut-off
dependence were different. The paper \cite{Rychkov:2015vap} used
a smaller Hilbert space of size $10000$ but decreased the truncation
errors drastically by deriving an effective Hamiltonian, which took
into account the cut-out Hilbert space in the leading non-vanishing
order of the perturbation theory. (This was further improved in \cite{Elias-Miro:2015bqk}).
Contrary, we used efficient diagonalizing routines for large sparse
matrices having up to $250$ million nonzero matrix elements, and
extrapolated in the cut dependence following their previous work \cite{Rychkov:2014eea}.
We tested this method in the symmetric sector by comparing our results
to the one in \cite{Rychkov:2014eea}: We used the code in \cite{Rychkov:2014eea}
to produce raw and renormalized data, which we compared to our raw
and extrapolated data. \textcolor{black}{At $g_{4}=1.5$, $m=1$,
using an energy cut} $e_{cut}=16$, the raw data agreed with high
precision: the difference appearing in the fourth-fifth digit for
the first eigenvalues (the truncation was slightly differently defined,
so deviations were larger than the numerical precision). Our extrapolated
data agreed with the renormalized data within their errors for all
observables, except for the ground-state energy density, for which
our data was a bit lower. Thus our method possibly slightly overextrapolates.
For this reason, in the future we are planning to combine the two
methods. 

In comparing the results we note that we managed to analyze the model
in a wider parameter range than \cite{Rychkov:2015vap}. For the overlapping
parameters we found agreement within the numerical errors. For large
couplings we identified the critical point $g=3.2\left(1\right)$,
which agrees with their findings using Chang duality. Since the numerical
errors grow with the volume we put emphasis on the finite size spectrum
including all polynomial and the leading exponential corrections.
We did it in all sectors of the theory. In measuring the kink mass
via the vacuum splitting we fitted their finite size formula. Having
a good control of the finite size effects enabled us to extract physical
information from the volume range $l=[5,15]$ where both the exponential
theoretical corrections and the numerical errors were small.

\section{Conclusions}

\begin{figure}[H]
\begin{centering}
\includegraphics[width=12cm]{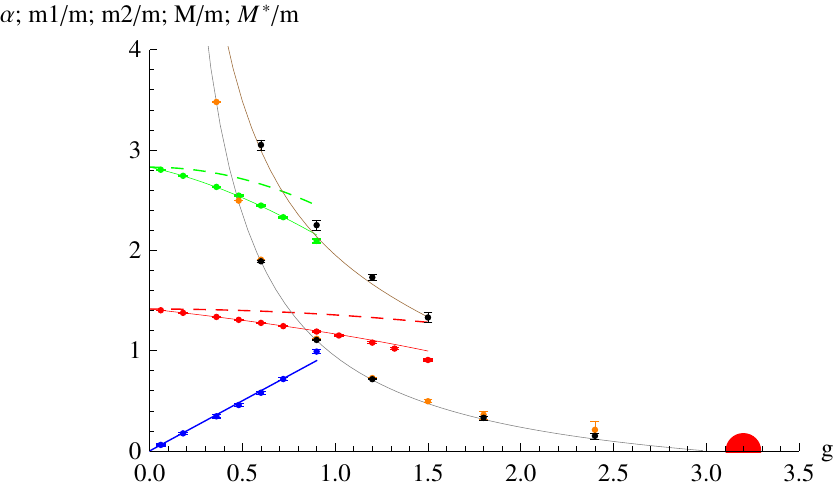}
\par\end{centering}

\protect\caption{The mass spectrum of the model as function of the coupling $g$. The
mass of the breathers, $m_{1}$ (red), $m_{2}$ (green), together
with the scattering parameter $\alpha$, (blue). The numerical $B_{2}$
mass is computed using eq. (\ref{eq:b2mass}). The semiclassical results
(\ref{eq:semimn},\ref{eq:semialpha}) are indicated by dashed lines.
The semiclassical result for the kink mass is shown in grey, while
the one determined from the vacuum splitting by black and the one
determined from the anti-periodical sector by orange. The semiclassical
excited kink is shown in brown. The perturbative correction to $m_{1}$
(\ref{eq:m1pert}) is indicated by a solid red line. The solid green
line depicts $m_{2}$ as calculated by (\ref{eq:b2mass}), using the
perturbative result for $m_{1}$ and the semiclassical value of $\alpha$.}

\centering{}\label{mass_spectrum}
\end{figure}

In this paper we investigated the two dimensional $\phi^{4}$ theory
in the broken phase by the massive generalization of the truncated
conformal space approach. We called this method the truncated Hilbert
space approach and we developed a code based on the diagonalization
of large sparse matrices to diagonalize the truncated dimensionless
Hamiltonian. We determined the spectrum numerically for various volumes
and quartic couplings and both for periodic and antiperiodic boundary
conditions. In the periodic case we found it useful to adapt the mini
superspace approach. We compared the results with semiclassical calculations
and also with finite size corrections. These finite size corrections
are expressed in terms of the infinite volume masses and scattering
matrices, for which we proposed parametrizations from the unitarity
relations. We summarized the infinite volume mass spectrum and scattering
parameter as functions of the coupling on Figure \ref{mass_spectrum}. 

We found a very good agreement with the semiclassical results for
the breather spectrum for moderate couplings. We then mapped the full
spectrum of the neutral and kink excitations. We found at most two
neutral excitations and each particle disappeared from the spectrum
when its mass exceeded twice the mass of the lightest particle. This
confirms the conjecture of \cite{Mussardo:2006iv} that only two neutral
excitations can exist in the spectrum. 

In determining the kink mass we used both the splitting of the even
and odd ground-state energies with periodical boundary conditions
and also the lowest lying state in the antiperiodic sector. These
two methods agreed very well and the results were close to the semiclassical
spectrum. For large couplings we identified the critical point beyond
which the broken $\mathbb{Z}_{2}$ symmetry is restored. We also confirmed
that the second order phase transition is within the Ising universality
class. In computing the leading exponential mass correction of the
lightest neutral excitation we found relevant deviations from the
Lüscher formula (\ref{eq:Fterm}). By comparing the spectrum with
the analogue breather spectrum of the sine-Gordon theory we got convinced
that this effect must be attributed to non-integrable, inelastic processes,
which need to be understood further. 

We believe that improving our data with the renormalization group
method of \cite{Rychkov:2014eea} enables to measure other interesting
quantities such as decay rates or the inelastic part of the scattering
matrices. 

Another interesting direction of research is to describe the strongly
coupled spectrum by perturbing the conformal Ising model and to compare
the analytical results to the improved numerics.

\section*{Acknowledgments}

We thank Zoltán Laczkó for a collaboration at an early stage. ZB thanks
Joao Penedones for the discussions on dispersion relations and the
parametrization of the scattering matrix, while ML would like to thank
László Ujfalusi for useful discussions regarding the used numerical
eigensolver package. We also thank Slava Rychkov and Lorenzo Vitale
for suggestions and for comments. This work was supported by a Lendület
Grant and by OTKA K116505.

\appendix

\section{Ground-state energy from normal ordering}

In this Appendix we motivate the ground-state energy 

\begin{equation}
E_{0}(\mu,L)=\mu\int_{-\infty}^{\infty}\frac{d\theta}{2\pi}\cosh\theta\log(1-e^{-mL\cosh\theta})
\end{equation}
of the free massive boson from a normal ordering prescription. The
free normal ordered Hamiltonian in volume $L$ takes the form
\begin{equation}
H^{\mu,L}=\frac{1}{2}\int_{0}^{L}:(\partial_{t}\phi)^{2}+(\partial_{x}\phi)^{2}+\mu^{2}\phi^{2}:_{\mu,L}
\end{equation}
Normal ordering is understood that we expand $\phi$ as 
\begin{equation}
\phi(x,t)=\sum_{n}\frac{1}{\sqrt{2\omega_{n}L}}\left(a_{n}e^{-i\omega_{n}t+ik_{n}x}+a_{n}^{+}e^{i\omega_{n}t-ik_{n}x}\right)
\end{equation}
with $\omega_{n}=\sqrt{\mu^{2}+k_{n}^{2}}$ and $k_{n}=\frac{2\pi n}{L}$
and then put all the annihilation operators $a_{n}$ on the right
of the creation operators $a_{n}^{+}$ using the rule
\begin{equation}
[a_{n},a_{m}]=[a_{n}^{+},a_{m}^{+}]=0\quad;\qquad[a_{n},a_{m}^{+}]=\delta_{n,m}
\end{equation}
This implies that 
\begin{eqnarray}
\phi^{2} & = & :\!\phi^{2}\!:_{\mu,L}+Z(\mu,L)\quad;\qquad Z(\mu,L)=\sum_{\vert n\vert<N}\frac{1}{2\omega_{n}L}\\
(\partial_{t}\phi)^{2} & = & :\!(\partial_{t}\phi)^{2}\!:_{\mu,L}+Y_{t}(\mu,L)\quad;\qquad Y_{t}(\mu,L)=\sum_{\vert n\vert<N}\frac{\omega_{n}^{2}}{2L\omega_{n}}\\
(\partial_{x}\phi)^{2} & = & :\!(\partial_{x}\phi)^{2}\!:_{\mu,L}+Y_{x}(\mu,L)\quad;\qquad Y_{x}(\mu,L)=\sum_{\vert n\vert<N}\frac{k_{n}^{2}}{2L\omega_{n}}
\end{eqnarray}
where we introduced a cut-off to make them finite. 

In the paper we are interested in the following perturbation of the
system
\begin{equation}
H=H^{\mu,L}+\frac{\delta\mu^{2}}{2}\int_{0}^{L}:\phi^{2}:_{\mu,L}dx=\frac{1}{2}\int_{0}^{L}:(\partial_{t}\phi)^{2}+(\partial_{x}\phi)^{2}+\mu'^{2}\phi^{2}:_{\mu,L}
\end{equation}
where $\mu'^{2}=\mu^{2}+\delta\mu^{2}$. We would like to understand
in which way it is related to $H^{\mu',L}$. They formally look the
same -- however, they are normal ordered at different mass scales.
As a result, they differ by a constant term. To calculate this constant,
we use
\begin{eqnarray}
:\!\phi^{2}\!:_{\mu,L} & = & :\!\phi^{2}\!:_{\mu',L}+\Delta Z\quad;\qquad\Delta Z=\frac{1}{2L}\sum_{n}\left(\frac{1}{\omega'_{n}}-\frac{1}{\omega_{n}}\right)\\
:\!(\partial\phi)^{2}\!:_{\mu,L} & = & :\!(\partial\phi)^{2}\!:_{\mu',L'}+\Delta Y\quad;\qquad\Delta Y=\frac{1}{2L}\sum_{n}\left(\frac{(\omega'_{n})^{2}+k_{n}^{2}}{\omega'_{n}}-\frac{\omega{}_{n}^{2}+k_{n}^{2}}{\omega{}_{n}}\right)\nonumber 
\end{eqnarray}
where $\omega'_{n}=\sqrt{\mu'^{2}+k_{n}^{2}}$ and $(\partial\phi)^{2}=(\partial_{t}\phi)^{2}+(\partial_{x}\phi)^{2}$.
Observe that all the sums are finite, although this would not be true
separately for $(\partial_{t}\phi)^{2}$ or for $(\partial_{x}\phi)^{2}$.
In the first term we can subtract the infinite volume limit
\begin{equation}
Z(\mu',\infty)-Z(\mu,\infty)=\int_{-\infty}^{+\infty}\frac{dk}{4\pi}\left(\frac{1}{\sqrt{\mu'^{2}+k^{2}}}-\frac{1}{\sqrt{\mu^{2}+k^{2}}}\right)=-\frac{1}{2\pi}\log\frac{\mu'}{\mu}
\end{equation}
to obtain
\begin{equation}
\Delta Z=Z(\mu',L)-Z(\mu,L)=z(\mu'L)-z(\mu L)-\frac{1}{2\pi}\log\frac{\mu'}{\mu}
\end{equation}
where 
\begin{equation}
z(\mu L)=Z(\mu,L)-Z(\mu,\infty)=\int_{0}^{\infty}\frac{d\theta}{\pi}(e^{\mu L\cosh\theta}-1)^{-1}
\end{equation}
Similarly, 
\begin{equation}
Y(\mu',\infty)-Y(\mu,\infty)=\int_{-\infty}^{+\infty}\frac{dk}{4\pi}\left(\frac{\mu'^{2}+2k^{2}}{\sqrt{\mu'^{2}+k^{2}}}-\frac{\mu^{2}+2k^{2}}{\sqrt{\mu^{2}+k^{2}}}\right)=\frac{1}{4\pi}\left(\mu'^{2}-\mu^{2}\right)
\end{equation}
so that we get
\begin{equation}
\Delta Y=\frac{2E_{0}(\mu',L)}{L}-\frac{2E_{0}(\mu,L)}{L}-\mu'^{2}z(\mu'L)+\mu^{2}z(\mu L)+\frac{1}{4\pi}\left(\mu'^{2}-\mu^{2}\right)
\end{equation}
This implies that 
\begin{equation}
H^{\mu,L}+\frac{\delta\mu^{2}}{2}\int_{0}^{L}:\phi^{2}:_{\mu,L}dx=H^{\mu',L}+E_{0}(\mu',L)-E_{0}(\mu,L)-\frac{\delta\mu^{2}}{2}Lz(\mu L)+L\left[\frac{1}{8\pi}(\mu'^{2}-\mu^{2})-\frac{\mu'^{2}}{4\pi}\log\frac{\mu'}{\mu}\right]
\end{equation}
We can reformulate this result as 
\begin{equation}
H^{\mu,L}+E_{0}(\mu,L)+LE_{bulk}(\mu)+\frac{\delta\mu^{2}}{2}\int_{0}^{L}\left(:\phi^{2}:_{\mu,\infty}-\frac{1}{2\pi}\log\mu\right)dx=H^{\mu',L}+E_{0}(\mu',L)+LE_{bulk}(\mu')
\end{equation}
That is, if we normal order the interaction term at infinite volume
and introduce the ground-state energy by $E_{0}(\mu,L)$ and ground-state
energy density by $E_{bulk}(\mu)=\frac{\mu^{2}}{8\pi}-\frac{\mu^{2}}{4\pi}\log\mu$
then the perturbation will change these parameters in a consistent
way. In the following, we do not keep track of the energy density
and introduce the perturbing operator simply as $:\phi^{2}:_{\mu,\infty}$. 

One can also wonder how the ground state energy changes by changing
the volume, i.e. to relate the system with mass $\mu$ and volume
$L$ to the same system with volume $L'$. Clearly the spectrum of
the two systems are completely different. What we can compare is the
ground state energy. We will demand that $LH$ depends only on $\mu L$.
This is satisfied for
\begin{equation}
LH^{\mu,L}+LE_{0}(\mu,L)=\frac{1}{2}\sum_{k}\sqrt{(\mu L)^{2}+4\pi^{2}n^{2}}+\mu L\int_{-\infty}^{\infty}\frac{d\theta}{2\pi}\cosh\theta\log(1-e^{-\mu L\cosh\theta})
\end{equation}
This implies that 
\begin{equation}
LH^{\mu,L}+LE_{0}(\mu,L)=L'H^{\mu',L'}+L'E_{0}(\mu',L')\quad;\qquad\mu L=\mu'L'
\end{equation}
By this rule we can calulate the ground state energy at volume $L'$
and mass $\mu$. 

Summarising, the vacuum energy at volume $L$ and renormalization
scale $\mu$ is $E_{0}(\mu,L)$, which is independent of the path
we arrive at this point, i.e. by adding perturbations or by changing
the volume. But we do not keep track of the change in the vacuum energy
density.

\section{Parametrizing the scattering matrix}

Here we analyze the $2\to2$ scattering matrices of the theory. We
first recall how the unitarity of the scattering matrix can be used
to gain information on its analytic structrure. The idea is to write
the unitarity equation $SS^{\dagger}=\mathbb{I}$ for the interaction
part of the matrix $S=\mathbb{I}+iT$ as
\begin{equation}
i(T-T^{\dagger})=-TT^{\dagger}
\end{equation}
Taking matrix element between initial and final states and inserting
a complete system we can write
\begin{equation}
\langle out\vert i(T-T^{\dagger})\vert in\rangle=-\sum_{n\in\mathcal{H}}\langle out\vert T\vert n\rangle\langle n\vert T^{\dagger}\vert in\rangle\label{eq:Tuni}
\end{equation}
For diagonal matrix elements, this equation implies the optical theorem,
namely the absolute square of the total cross section is related to
the imaginary part of the forward scattering amplitude. This equation
shows that S-matrix elements are not independent, they are connected
analytically by a very intrinsic way. 

Let us focus on the $2\to2$ particle scattering process and denote
the momenta of the initial particles as $p_{1},p_{2}$ and outgoing
particles as $p_{3},p_{4}$ , which are all onshell $p_{i}^{2}=\omega_{i}^{2}-k_{i}^{2}=m_{1}^{2}$.
Energy conservation can be factored out 
\begin{equation}
\langle p_{3},p_{4}\vert S\vert p_{1},p_{2}\rangle\propto(2\pi)^{2}\delta^{(2)}(p_{1}+p_{2}-p_{3}-p_{4})S(s,t,u)
\end{equation}
where due to relativistic invariance the scattering matrix depends
on the Mandelstam variables
\begin{equation}
s=(p_{1}+p_{2})^{2}\quad;\qquad t=(p_{1}-p_{3})^{2}\quad;\qquad u=(p_{1}-p_{4})^{2}
\end{equation}
On-shell condition implies that $s+t+u=4m_{1}^{2}$. In two dimension
the kinematics is further restricted since $t=4m^{2}-s$. The analytical
S-matrix theory states that $S(s)$ can have poles and cuts dictated
by unitarity%
\footnote{There can also be anomalous thresholds corresponding to diagrams with
onshell propagating particles, but we assume that they are not relevant
here.%
}, but otherwise it is an analytical function of $s$, which also satisfies
crossing symmetry. Spelling out the equation (\ref{eq:Tuni}) for
the two-particle process we obtain:
\begin{equation}
\langle p_{3},p_{4}\vert i(T-T^{\dagger})\vert p_{1},p_{2}\rangle=-\sum_{n\in\mathcal{H}}\langle p_{3},p_{4}\vert T\vert n\rangle\langle n\vert T^{\dagger}\vert p_{1},p_{2}\rangle
\end{equation}
In factoring out momentum coservation we assume that $k_{1}>k_{2}$
and $k_{3}>k_{4}$ such that $k_{1}=k_{3}$ and $k_{2}=k_{4}$ must
hold. Let us write out the first few terms. Clearly the vacuum does
not contribute. The one- and two-particle contributions read as 
\begin{eqnarray}
2\Re eS(s,t) & = & \int dk_{n}\delta^{(2)}(p_{1}+p_{2}-p_{n})\Gamma_{12}^{n}\Gamma_{n}^{34}+\\
 &  & \int dk_{n_{1}}dk_{n_{2}}\delta^{(2)}(p_{1}+p_{2}-p_{n_{1}}-p_{n_{2}})S(p_{1},p_{2},p_{n_{1}},p_{n_{2}})S(p_{n_{1}},p_{n_{2}},p_{3},p_{4})+\dots\nonumber 
\end{eqnarray}
Here we factored out the total energy momentum conservation. However,
on the rhs. other $\delta$ functions appeared, which eat up one integration
leading to an energy delta function in the one-particle and a constant
piece in the two-particle contribution. Since these contributions
appear only for certain analytically continued values of $s$, they
tell us the non-analytical positions of the scattering matrix on the
complex $s$-plane. Bound states correspond to poles, where energy
conservation can be maintained, and the two-particle threshold gives
a cut starting from $s=4m_{1}^{2}$. 

Let us go into the center of mass frame $k_{1}\equiv k=-k_{2}$. Then
we have $s=4m_{1}^{2}+4k^{2}$ and $t=-4k^{2}$, such that relation
$t=4m_{1}^{2}-s$ is satisfied. Crossing symmetry implies that 
\begin{equation}
S(s+i\epsilon)=S(t+i\epsilon)=S(4m_{1}^{2}-s-i\epsilon)
\end{equation}
Switching from the $i\epsilon$ prescription to the $-i\epsilon$
prescription is equivalent to time reversal, which changes the scattering
matrix to its inverse 
\begin{equation}
S(s+i\epsilon)=S(s-i\epsilon)^{-1}
\end{equation}
 The analytical structure of the S-matrix is displayed on the following
figure: 

\begin{center}
\includegraphics[width=12cm]{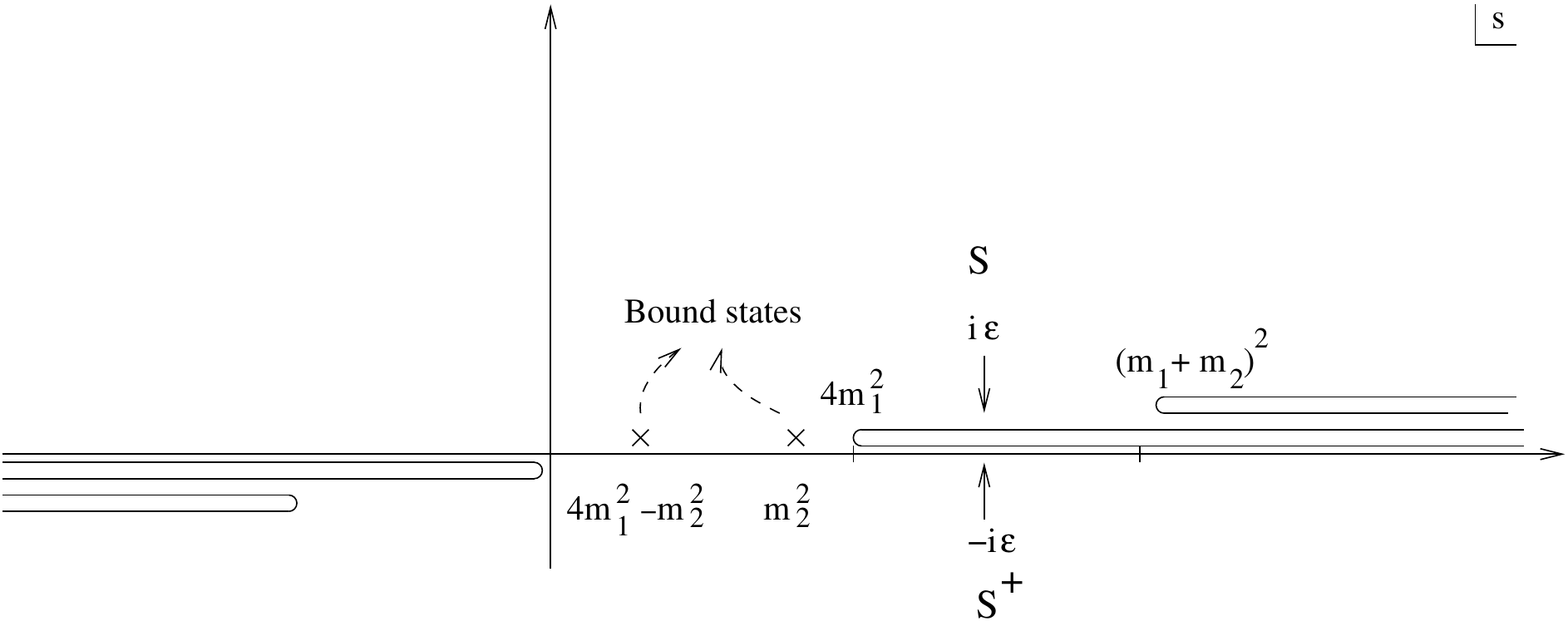}
\par\end{center}

The bound state poles are at $m_{2}^{2}$ and at $4m_{1}^{2}-m_{2}^{2}$.
There are cuts on the real line starting at the multiparticle thresholds
$(m_{n}+m_{k}+\dots)^{2}$. The physical value of the S-matrix is
at $S(s+i\epsilon)$ where $s>4m_{1}^{2}$.

In the following we parametrize the two-particle scattering matrix
based on this analytical structure. We first switch to rapidity parametrization.
In the center of mass frame $k_{1}=-k_{2}=m\sinh\frac{\theta}{2}$,
such that the rapidity difference is $\theta$. The rapidity is related
to the $s$ variable as 
\begin{equation}
s=4m_{1}^{2}(1+\sinh^{2}\frac{\theta}{2})=4m_{1}^{2}\cosh^{2}\frac{\theta}{2}
\end{equation}
This resolves the cut starting at $4m_{1}^{2}$ and maps the first
sheet of the complex $s$-plane to the strip $0<\Im m(\theta)<\pi$.
The bound state poles are located on the imaginary axes at $\theta=i\zeta$
and at $\theta=i(\pi-\zeta)$, since 
\begin{equation}
m_{2}^{2}=4m_{1}^{2}\cos^{2}\frac{\zeta}{2}
\end{equation}
The first branch cut depends on the masses. Assuming the kinks are
very heavy they appear at $s=(m_{1}+m_{2})^{2}$ or in the rapditity
parameter at $\theta_{t}$: 
\begin{equation}
\cosh\frac{\theta_{t}}{2}=\frac{1}{2}(1+\frac{m_{2}}{m_{1}})=\frac{1}{2}\bigl(1+2\cos\frac{\zeta}{2}\bigr)
\end{equation}
If the kinks are lighter, then even earlier. Now we turn this information
into a parametrization of the scattering matrix. Recall that crossing
symmetry translates to the rapidity parameter as 
\begin{equation}
S(i\pi-\theta)=S(\theta)
\end{equation}
while the S-matrices on the two sides of the cut are related as
\begin{equation}
S(\theta)=S(-\theta)^{-1}.
\end{equation}
In order to have the right periodicity and good asymptotic behaviour
let us consider the logarithmic derivative of the scattering matrix:
\begin{equation}
\phi(\theta)=\frac{d}{d\theta}\log S(\theta)
\end{equation}
It has the properties
\begin{equation}
\phi(\theta)=\phi(-\theta)\quad;\qquad\phi(\theta)=-\phi(i\pi-\theta)=-\phi(i\pi+\theta)
\end{equation}
Let us write it as
\begin{equation}
\phi(\theta)=\oint_{\theta}\frac{d\theta'}{2\pi i}\frac{\phi(\theta')}{\sinh(\theta'-\theta)}
\end{equation}
where the contour sorrounds infinitesimally $\theta$. Now we blow
up the contour and deform it to $\theta=0$ above the cuts and to
$\theta=i\pi$ below the cuts. In doing so we pick up the residues
of the bound state poles: 
\begin{equation}
\phi(\theta)=\frac{1}{\sinh(i\zeta-\theta))}+\frac{1}{\sinh(i(\pi-\zeta)-\theta)}-\int_{\theta_{t}}^{\infty}\frac{d\theta'}{2\pi i}\rho(\theta')\left(\frac{1}{\sinh(\theta'-\theta)}+\frac{1}{\sinh(i\pi-\theta'-\theta)}\right)
\end{equation}
where $\rho(\theta)$ is the jump of $\phi(\theta)$ on the cut, which
is related to the inelastic scattering processes through the unitarity
relation. After integrating and exponentiating we obtain
\begin{equation}
S(\theta)=S(\theta,i\zeta)\exp\left\{ -\int_{\theta_{t}}^{\infty}\frac{d\theta'}{2\pi i}\rho(\theta')\log S(\theta',\theta)\right\} \quad;\qquad S(\theta,x)=\frac{\sinh\theta+\sinh x}{\sinh\theta-\sinh x}
\end{equation}
where we made the integral to converge for large $\theta'$. We note
that similar parametrization should be valid for the kink-kink and
kink-breather scattering matrices.

\section{A review of the numerical algorithm used}

In this Appendix we review the numerical implementation of our method.

\subsection{Overview}

The numerical results were obtained using an application written in
C++, that is supplied with this document. We tested the simulation
under \emph{SuSe} and \emph{Ubuntu Linux} platforms.

As explained earlier,the matrix elements of the Hamiltonian are calculated
in a truncated, finite-dimensional basis and then the lowest eigenvalues
of the obtained large sparse matrix is determined. With the use of
the PRIMME iterative eigensolver package\cite{PRIMME} (of which we
currently use version 1.1), it was possible to work with matrices
of up to $250$ million nonzero mattix elements on a single personal
computer, the computing time being a couple of minutes per measurement
point. The eigensolver uses an improved version of the Jacobi-Davidson
method (referred as \emph{JDQMR\_ETOL} in PRIMME).

The algorithm itself is generally capable of handling a Hamiltonian
composed of the sum of a finite number of Landau-Ginzburg interactions
in 1+1 dimensions put into a finite volume. In this article, we focused
on the parity-conserving $\varphi^{4}$ model, and thus the parameter
corrections are only implemented to this case. In the periodic sector,
it proved to be sufficient to use the minimal Hilbert-space approach,
with the separate treatment of the zero mode and the construcion of
the overall Hamiltonian as a direct product of the minimal and non-minimal
subspaces. An emphasis was also put on the antiperiodic sector, though
obviously the minimal space approach is not applicable in this case. 

A single run of the program usually generates several output files,
each one corresponding to a different energy cutoff. A single file
contains the relevant eigenvalues for different volumes between a
minimal volume $L_{min}$ and maximal volume $L_{max}$, having step
$L_{step}$, the interaction strength parameters being corrected corresponding
to the change of volume, while all other parameters are kept fixed.

The parameters of the simulations can be set in a separate configuration
file, in which the user can set the number of eigenvalues to be calculated,
along with the interaction strengths, truncation dimensions, and output
file name. It is possible there to decide the usage of the minimal
space method, choose the boundary condition and select the desired
parity and momentum sector. A basis mass and volume must also be given
for the construction of the free Fock basis. The examined volume interval
and step size are also to be given in the aforementioned file. 

The source code is structured around C++ classes and is divided into
several files. The entry point of the program is in the file \emph{main.cpp}.
Some auxiliary routines and numerical functions are declared and defined
in \emph{utils.h} and \emph{utils.cpp}. A numerical integration routine
from \emph{Numerical Recipes\cite{Press:1988:NRC:42249}} is implemented
in \emph{numint.cpp}. The classes \emph{HilbertState} and \emph{HilbertSpace},
which are used to build the Fock basis, are defined in \emph{phi4.cpp}.
The algorithms regarding the calculation of a Landau-Ginzburg interaction
term between two Fock states are implemented in \emph{phi\_n.cpp}.
The class \emph{Hamiltonian} is realized in \emph{minspace.h} and
\emph{minspace.cpp}. A class to store a sparse matrix in coordinate
list (COO) format is defined in \emph{coo.cpp}, along with a number
of routines particularly useful for our considerations. The header\emph{
primme.h} provides the interface to PRIMME functions. The connection
between the simulation and the eigensolver is realised in the utility
functions of \emph{PRIMME\_int.h.} Most notably, as PRIMME uses a
different sparse matrix format (Compressed Sparse Row, CSR), there
is a routine to convert the COO format sparse matrix to one that is
operable by the eigensolver.

\subsection{The basis}

A state of the Fock space is represented by the class HilbertState
in the algorithm. In fact, the particle content can be written in
conventional representation
\[
\left|u\right\rangle =\left|\left(k_{1},n_{1}\right),\left(k_{2},n_{2}\right),\dots\left(k_{r},n_{r}\right)\right\rangle 
\]
where this notation stands for having a state $\left|u\right\rangle $
that contains a number $n_{1}$ of particles with momentum $k_{1}$,
and a number $n_{2}$ of particles with momentum $k_{2}$, and so
on, until all particles are counted for in the state. Due to the finite
volume, the momenta are quantized so that in the periodic sector:
\[
k_{m}=\frac{\pi}{L}\left(2m\right),\quad m\in\mathbb{Z},
\]
while in the antiperiodic sector,
\[
k_{m}=\frac{\pi}{L}\left(2m+1\right),\quad m\in\mathbb{Z}_{-};\qquad k_{m}=\frac{\pi}{L}\left(2m-1\right).\quad m\in\mathbb{Z}_{+}
\]
Any Fock vector is thus identified by a finite series of integer pairs.
A series like this can be stored in a variable of type \emph{HilbertState}.

The \emph{HilbertSpace} class contains an array of \emph{HilbertState}
variables. Crucially, this array is filled by a member function in
a determined way, so that the \emph{HilbertSpace} contains all Fock
vectors below a given energy cutoff. Starting from an auxiliary vacuum
state (a \emph{HilbertState} variable that is empty), the construction
is performed recursively, each recursive step having two parts: first
we add an additional particle of some momentum $i$ to the auxiliary
Fock state containing only particles with positive momenta. Then we
distribute the signs of particle momenta every possible way so that
the overall momentum lies in the appropriate sector. The construction
routine is called with a starting auxiliary state, and has a loop
in which the function calls itself with another auxilary state having
one additional particle. If the energy cut is reached, the loop breaks
and the routine returns. If the embedded routine returns, the momentum
i of the added particle gets increased. At the end, the members of
the array get sorted according to their energy (with respect to the
basis volume and mass).

The truncation of basis is also implemented in the \emph{HilbertSpace}
class; it corresponds to cutting off the highest energy states from
the array.

\subsection{Calculating interaction matrix elements}

Due to the remarkable efficiency of the eigensolver routine, the construction
of the sparse matrix and evaluation of matrix elements can also turn
out to be the bottleneck of computation speed. The algorithm was designed
to fit two criteria: firstly it should be able to decide very quickly
if a matrix element is zero or not, and secondly, it should calculate
the actual values of matrix elements as effectively as possible.

A Landau-Ginzburg interaction term of degree $N$ reads
\begin{equation}
\mathcal{O}_{N}=g_{N}\intop_{0}^{L}:\varphi\left(x,t\right)^{N}:dx=g_{N}\frac{1}{2^{N/2}L^{N/2-1}}\sum_{\left\{ k_{i},\epsilon_{i}\right\} }\frac{:\prod_{i=1}^{N}a_{k_{i}}^{\epsilon_{i}}:}{\sqrt{\prod_{i=1}^{N}\omega_{n_{i}}}}\left(\delta_{\sum_{i=1}^{N}\epsilon_{i}k_{i}}\right)\label{eq:LGN}
\end{equation}
where $k_{i}$ runs through the one-particle momenta allowed by the
boundary condition, $\epsilon=\pm1$, $a_{n}^{+1}\equiv a_{n}^{+}$,
and $a_{n}^{-1}\equiv a_{n}$.

First it should be noted that the matrix element $\left\langle a\right|\mathcal{O}_{N}\left|b\right\rangle $
can have a nonzero value only if the state $\left|b\right\rangle $
can be transformed into a state equivalent of $\left|a\right\rangle $
by a product of at most $N$ creation-annihilation opeators. Therefore
the problem of fast discrimination of zero matrix elements reduces
to using the particle content of $\left|a\right\rangle $ to generate
all states $\left|b\right\rangle $ with energy $E_{\left|b\right\rangle }\leq E_{\left|a\right\rangle }$
and $\left\langle a\right|\mathcal{O}_{N}\left|b\right\rangle \neq0$
and looking them up in the array of \emph{HilbertState}s. Secondly,
the actual content difference of $\left|a\right\rangle $ and $\left|b\right\rangle $
determines which operator product terms from the sum in \eqref{eq:LGN}
contribute to the matrix element. (Since the simulation imposes normal
ordering at the volume of computation, we don't have to consider terms
of ``tadpole'' type, where new particles not contained in any of
the states would pop up and annihilate each other.)

The determination of the content difference is facilitated by the
class \emph{StateDifference}. This structure contains two arrays:
the first array lists the momentum of all particles in the bra vector
that are not present in the ket vector; the secod contains the particles
only present in the ket vector. The sum of array sizes gives the total
number of creation-annihilation operators needed to transform the
states into each other. This total number will be called particle
difference for simplicity. The comparison of the two \emph{HilbertSpace}
variables and the construction of these arrays is done by a constructor
of \emph{StateDifference}.

If the particle difference of two \emph{HilbertState}s is exactly
$N$, then there is only one type of contributing normal ordered operator
product from \eqref{eq:LGN}. However, if we only count operator product
terms that are different after normal ordering, we always have to
remember that the action of normal ordering effectively produces a
multinomial coefficient for each term. If the particle difference
is a value $N_{pd}<N$, then we may also annihilate and create $N-N_{pd}$
particles taken arbitarily from the ket vector. Since the matrix is
symmetric, we can interchange the bra and ket vector so that the number
of annihilation operators is always greater or equal than the number
of creation operators in the term under consideration. Thus every
contributing term can be accounted for by examining the ket vector
particle content and the particle difference. The contributions to
the matrix element of $\mathcal{O}_{n}$ are then calculated term
by term. This computation is done for a single $\mathcal{O}_{N}$
term using the class \emph{phi\_n\_interaction}.

\subsection{The \emph{Hamiltonian}}

As mentioned earlier, the matrix representation of the truncated Hamiltonian
in the generated Fock basis is stored in a sparse matrix. The matrix
is technically stored in a variable of type \emph{sparse\_matrix}.
A single matrix element is kept in a struct variable such that the
row and column index is stored explicitly along with the value of
the element. Tha class \emph{sparse\_matrix} consists of an array
of these structs, along with some useful member functions.

The interactive part of the Hamiltonian consists of a sum of terms
$\mathcal{O}_{N}$ in the form \eqref{eq:LGN}. The parameter corrections
according to the scheme change are accounted for in the main function
of the program. 

The Hamilton operator is technically realized by a class \emph{Hamiltonian}
in the program. If the minimal space approach is not used, then at
the beginning of the calculation, the matrix is filled with the matrix
elements of the interaction, the non-diagonal ones being only saved
if they are nonzero. In the Fock basis, the free Hamiltonian is diagonal,
and its matrix elements are also added to the matrix. After this initial
construction, only the nonzero matrix elements are modified.

\subsubsection{Minimal space approach}

If this approach is used, the program initially calculates the normalized
(i.e. $g_{N}=1$) matrix elements of $\mathcal{O}_{N}$ between non-minimal
basis states for every positive $N$ smaller or equal than the maximal
interaction exponent appearing in the Hamilton operator, and store
the representations in an array of sparse matrices contained in the
class \emph{Hamiltonian}.

For each examined volume, the minimal space spanned by zero modes
are accounted for first. This basically corresponds to solving an
anharmonic oscillator with appropriate coefficients and obtaining
the first eigenvalues and eigenvectors in the free basis. The $\varphi^{0}$
terms of the interaction are accounted for in the minimal space calculation.
As a second step, the matrix elements of interaction terms $\mathcal{O}_{N}$
are calculated between the numerically determined ``exact'' eigenvectors
in the minimal space, and get stored as an array of (dense) matrices.
Then the non-minimal (normalized) matrix elements of every $\mathcal{O}_{N}$
get actualised in the stored array of sparse matrices. In the last
step, the representation of the truncated Hamilton-operator is obtained
as a sum of direct product terms according to (\ref{eq:hfull}).

If the computation is limited to a single parity sector, then the
zero-mode oscillator is calculated separately in the even and odd
particle number basis. The minimal-space $\mathcal{O}_{N}$ matrices
in the exact basis are constructed so that the basis vectors containing
an even number of particles are ahead of those having an odd number
of particles. Then the non-minimal representations of $\mathcal{O}_{N}$s
are also transformed so that the even basis elements are ahead of
the odd basis elements. Then the overall Hamiltonian is built up in
the desired sector using the appropriate submatrices.

\providecommand{\href}[2]{#2}\begingroup\raggedright\endgroup

\end{document}